\documentclass[seceq]{ptptex}
\usepackage{amsmath}
\usepackage{amsfonts}

\begin{document}
\baselineskip=7mm
\renewcommand{\baselinestretch}{1.28}
\newcommand{\ba}{\begin{eqnarray}}
\newcommand{\ea}{\end{eqnarray}}
\newcommand{\non}{\nonumber\\}


\pagestyle{plain}

\begin{flushright}
OCHA-PP-275 \\
RUP-07-1
\end{flushright}

\hskip 12.36cm{}
\vspace{3cm}

\centerline{\LARGE  General Operator Solutions and BRST Quantization}
\vskip 2mm
\centerline{\LARGE of Superstrings in the {\it pp}-Wave with Torsion}

\vspace{1.5cm}

\centerline{\bf Yoichi Chizaki$^{1,}$\!\!
\footnote{E-mail: {\tt yoichizaki@phys.ocha.ac.jp}}
and Shigeaki Yahikozawa$^{2,}$\!\!
\footnote{
E-mail: {\tt yahiko@rikkyo.ac.jp}}}
\vspace{1mm}

\centerline{\it $^1$Department of Physics, Ochanomizu University,
Tokyo 112-8610, Japan}

\centerline{\it $^2$Department of Physics, Rikkyo University,
Tokyo 171-8501, Japan}

\vspace{1cm}
 
\centerline{\bf Abstract}
\vspace{2mm}

We completely accomplish the canonically covariant quantization of
Ramond-Neveu-Schwarz (RNS) superstrings in the {\it pp}-wave 
background with a non-zero flux of the NS-NS antisymmetric 
two-form field.
Here this flux is equivalent to a nonvanishing torsion.
In this quantization, general operator solutions,
which satisfy the entire equation of motion and all the canonical
(anti)commutation relations,
play an important role.
The whole of covariant string coordinates and fermions can be 
composed of free modes.
Moreover, employing covariant free-mode representations,
we calculate the anomaly in the super-Virasoro algebra 
and determine the number of dimensions of spacetime and 
the ordering constant from the nilpotency condition 
of the BRST charge in the {\it pp}-wave background with the flux.


\newpage
\section{Introduction}
The perfect quantization of superstrings in a variety of
backgrounds, for example curved spacetime, is of great importance. 
In particular, the canonical quantization is significant for searching 
the fundamental laws of physics, including the string landscape, 
the string field theory, the AdS/CFT correspondence and matrix models.
Of course, quantization in a number of interesting backgrounds 
has already been investigated and has been applied to many models.
However, the canonically covariant quantization,
which is the point to understand the fundamental structure,
of superstrings in the curved spacetime background 
is not yet accomplished.
Therefore, it is necessary to do new development,
and it is effective to make a thorough investigation of
the canonically covariant quantization for superstrings in the simple
curved spacetime background.

The purpose of this paper is to canonically quantize the 
Ramond-Neveu-Schwarz (RNS) superstring in the {\it pp}-wave 
background with a non-zero flux of the NS-NS antisymmetric 
two-form field, by using the covariant BRST operator formalism.
The method of the covariant BRST quantization of bosonic strings 
in the {\it pp}-wave background with the flux,
which we have already performed\cite{Chi-Ya},
is very useful for canonical quantization of the superstrings.
Moreover, it is particularly important to construct general operator 
solutions and covariant free-mode representations.
First, we construct the general operator solutions which satisfy 
the entire equation of motion and all the canonical 
(anti)commutation relations.
Here the equations of motion mean the Heisenberg equations 
of motion whose form is that of the Euler-Lagrange equations 
of motion in the {\it pp}-wave background with the flux.
The general operator solutions of covariant string coordinates
and fermions can be composed in the covariant free-mode 
representations.
Second, by using the free-mode representations of
the covariant string coordinates and fermions 
for the energy-momentum tensor and the supercurrent, 
we calculate the anomaly in the super-Virasoro algebra 
and determine the number of dimensions of spacetime 
and the ordering constant from the nilpotency condition of 
the BRST charge in the {\it pp}-wave background with the flux.
Our covariant quantization has a deep meaning 
and will bring a new effect,
though there are other methods,
\cite{Horowitz-Steif-1}\tocite{Kunitomo}
for example, the light-cone quantization 
and the conformal field theory.

This paper is organized as follows.
In \S2 we briefly review the action of the RNS superstring in general
background fields and the {\it pp}-wave background with the flux.
In \S3 we obtain the equations of motion of a closed
superstring in the {\it pp}-wave background with the flux
and construct the general solutions of them.
In \S4 we make Dirac brackets in the general background
by using the second-class constrains for Majorana fermions,
and we complete all the canonical (anti)commutation relations.
We also explain the canonical anticommutation relations
of the Ramond (R) fermions and the Neveu-Schwarz (NS) fermions.
It is important to note the anti-periodic delta function 
appearing in the NS sector.
In \S5 we present new free-mode representations of 
all the covariant string coordinates and fermions, 
and we write up the definition of the normal orderings about
the modes of twisted fields.
The general operator solutions are explained in \S4 and \S5.
In \S6 we prove that both the general operator solutions and
the free-mode representations satisfy
all the equal-time canonical (anti)commutation relations among
all the covariant string coordinates and fermions.
In \S7 we calculate the anomaly in the super-Virasoro algebra 
by using the energy-momentum tensor and the supercurrent
in the free-mode representations of
all the covariant string coordinates and fermions.
We also comment on the super-Virasoro algebra of ghosts and
antighosts.
In \S8 we determine the number of dimensions of spacetime 
and the ordering constant from the nilpotency condition of 
the BRST charge in the {\it pp}-wave background with the flux.
Section 9 contains some conclusions.


\section{Action and Backgrounds}

Before discussing our models, we explain our notation.
We take as the flat world-sheet metric $\eta^{ab}={\rm diag}(-1,+1)$
and choose the representation of the two dimensional Dirac matrices
$\gamma^{0}=i\sigma^{2},\ \gamma^{1}=\sigma^{1}$. 
Moreover we define $\gamma_{3}=\gamma^{0}\gamma^{1}=\sigma^{3}$. 
Here $\sigma^{i} \ (i=1,2,3)$ are Pauli matrices. 
The Dirac matrices obey the Dirac algebra
$\{\gamma^{a},\gamma^{b}\}=2\eta^{ab}$
and satisfy the following relation 
$\gamma^{a}\gamma^{b}=\eta^{ab}+\epsilon^{ab}\gamma_{3}$,
where $\epsilon^{ab}$ is the totally world-sheet anti-symmetric 
tensor ($\epsilon^{01}=+1$).
We define a two dimensional Majorana spinor 
$\psi=\begin{pmatrix}\psi_{+}\\ \psi_{-}\end{pmatrix}$, 
and we use  the components of the Majorana spinor as $\psi_{A}\ (A=+,-)$. 
The Dirac conjugate of $\psi$ is 
$\bar{\psi}=\psi^{\dagger}\gamma^{0}=\psi^{\rm T}\gamma^{0}=(-\psi_{-},\psi_{+})$.

We consider the RNS superstring in the background fields which are
a general curved spacetime metric $G_{\mu\nu}$ 
and a NS-NS anti-symmetric tensor field $B_{\mu\nu}$.
Our starting point is the total action 
$S=S_{\rm M}+S_{\rm GF+gh}$, where $S_{\rm M}$ has the local supersymmetry 
and the total action $S$ is BRST invariant. 
The action $S_{\rm M}$ with the string coordinates, the Majorana fermions, 
the zweibein and the world-sheet gravitino is\cite{C-Z}\tocite{D-M-R-2}
\begin{align}
S_{\rm M}=-&\frac{1}{4\pi\alpha'}\int d\tau d\sigma 
e \Big[(g^{\alpha\beta}G_{\mu\nu}+e^{-1}\epsilon^{\alpha\beta}
B_{\mu\nu})\partial_{\alpha}X^{\mu}\partial_{\beta}X^{\nu}
+iG_{\mu\nu}\bar{\psi}^{\mu}\gamma^{\alpha}{\cal D}_{\alpha}\psi^{\nu}
\nonumber
\\
&+\frac{1}{8}{\cal R}_{\mu\rho\nu\sigma}\bar{\psi}^{\mu}(1+\gamma_{3})
\psi^{\nu}\bar{\psi}^{\rho}(1+\gamma_{3})\psi^{\sigma}
\nonumber\\
&+2iG_{\mu\nu}\bar{\chi}_{\alpha}\gamma^{\beta}\gamma^{\alpha}\psi^{\mu}
\partial_{\beta}X^{\nu}+\frac{1}{2}G_{\mu\nu}\bar{\chi}_{\alpha}\gamma^{\beta}
\gamma^{\alpha}\chi_{\beta}\bar{\psi}^{\mu}\psi^{\nu}
+\frac{1}{6}H_{\mu\nu\rho}\bar{\chi}_{\alpha}
\gamma^{\beta}\gamma^{\alpha}\psi^{\mu}\bar{\psi}^{\nu}
\gamma_{\beta}\gamma_{3}\psi^{\rho}\Big]\,.
\label{Action-M}
\end{align}
The action $S_{\rm GF+gh}$ with ghosts and antighosts is
\begin{align}
S_{\rm GF+gh}=\frac{1}{2\pi}
\int d\tau d\sigma e \left[
{\cal B}_{1\alpha}^{\ a}(e^{\alpha}_{\ a}-\delta^{\alpha}_{\ a})
+\bar{\cal B}^{\alpha}_{2}\chi_{\alpha}
-i b_{\alpha\beta}\nabla^{\alpha}c^{\beta}
+\bar{\beta}_{\alpha}\nabla^{\alpha}\gamma \right]\,,
\end{align}
where the action includes the gauge-fixing condition
and the antighosts are restricted from the conditions 
of the Weyl symmetry, the super-Weyl symmetry and the local Lorentz
symmetry.
We have to note that the indices $a,b$ denote the local Lorentz world-sheet 
indices which run over 0,1, the greek indices $\alpha,\beta$ denote 
the curved world sheet indices which run over 0,1 and the greek indices 
$\mu,\nu,\rho,\sigma,\lambda,\omega$ denote the spacetime indices 
which run over $0,1,\cdots,D-1$. 
The fields which appear on this world-sheet are string coordinates 
(world-sheet bosonic fields) $X^{\mu}$ and their superpartners 
(world-sheet Majorana fermionic fields) $\psi^{\mu}_{A}$, zweibein 
$e_{\alpha}^{\ a}$ and their fermionic superpartners (gravitino) 
$\chi_{\alpha}$. 
We define the curved world-sheet metric  $g_{\alpha\beta}$
and the determinant of the zweibein as follows: 
$g_{\alpha\beta}=e_{\alpha}^{\ a}e_{\beta}^{\ b}\eta_{ab},\ 
e=\det e_{\alpha}^{\ a}$.
Moreover we had better define the inverse of the zweibein as
$e^{\alpha}_{\ a}=(e_{\alpha}^{\ a})^{-1}$ for useful. 
Then Dirac matrices in curved world-sheets become 
$\gamma^{\alpha}=e^{\alpha}_{\ a}\gamma^{a}$. 
Here we define the covariant derivative ${\cal D}_{\alpha}$ 
and the generalized Riemann tensor ${\cal R_{\mu\rho\nu\sigma}}$, 
which contain the flux of the NS-NS anti-symmetric field, 
as follows:
\begin{align}
{\cal D}_{\alpha}\psi^{\nu}&=\partial_{\alpha}\psi^{\nu}
+\left(\Gamma^{\nu}_{\rho\sigma}+\frac{1}{2}\gamma_{3}H^{\nu}_{\rho\sigma}
\right)\partial_{\alpha}X^{\rho}\psi^{\sigma},\\
{\cal R}_{\mu\rho\nu\sigma}&=R_{\mu\rho\nu\sigma}
+\frac{1}{2}\left(\nabla_{\nu}H_{\mu\rho\sigma}-\nabla_{\sigma}H_{\mu\rho\nu}
\right)
+\frac{1}{4}G_{\lambda\omega}\left(H^{\lambda}_{\rho\nu}H^{\omega}_{\mu\sigma}
-H^{\lambda}_{\rho\sigma}H^{\omega}_{\mu\nu}\right)
\end{align}
We use the Christoffel symbol $\Gamma^{\mu}_{\rho\sigma}$, 
the field strength of 
the NS-NS anti-symmetric field $H_{\mu\nu\rho}$, the ordinary Riemann tensor 
$R_{\mu\nu\rho\sigma}$ and the spacetime covariant derivative of 
the NS-NS flux $\nabla_{\mu}H_{\nu\rho\sigma}$,
whose definitions are as follows:
\begin{align}
\Gamma^{\mu}_{\rho\sigma}&=\frac{1}{2}G^{\mu\nu}
\left(\partial_{\sigma}G_{\nu\rho}+\partial_{\rho} G_{\nu\sigma}
-\partial_{\nu}G_{\rho\sigma}\right)\,,\\
H_{\mu\nu\rho}&=\partial_{\mu}B_{\nu\rho}+\partial_{\nu}B_{\rho\mu}
+\partial_{\rho}B_{\mu\nu}\,,\\
R_{\mu\nu\rho\sigma}&=G_{\mu\lambda}(\partial_{\rho}
\Gamma^{\lambda}_{\nu\sigma}-\partial_{\sigma}\Gamma^{\lambda}_{\nu\rho}
+\Gamma^{\lambda}_{\omega\rho}\Gamma^{\omega}_{\nu\sigma}
-\Gamma^{\lambda}_{\omega\sigma}\Gamma^{\omega}_{\nu\rho})\,,\\
\nabla_{\mu}H_{\nu\rho\sigma}&=\partial_{\mu}H_{\nu\rho\sigma}
-\Gamma^{\lambda}_{\nu\mu}H_{\lambda\rho\sigma}
-\Gamma^{\lambda}_{\rho\mu}H_{\nu\lambda\sigma}
-\Gamma^{\lambda}_{\sigma\mu}H_{\nu\rho\lambda}\,.
\end{align}
In the action $S_{\rm GF+gh} =S_{\rm GF}+S_{\rm gh}$,
$S_{\rm GF}$ is the gauge fixing action and
$S_{\rm gh}$ is the Fadeev-Popov ghost action. ${\cal B}^{\ a}_{1 \alpha}$ and
${\cal B}^{\alpha}_{2}$ are the auxiliary fields to fix the gauge. $c^{\alpha}$, 
$b_{\alpha\beta}$ and $\gamma$, $\beta_{\alpha}$ are, respectively, 
the ghost field, the antighost field and their superpartners.
Since we achieve the covariant BRST quantization for string theory in this paper, 
we choose the covariant gauge-fixing condition on the world-sheet; 
$e^{\alpha}_{\ a}=\delta^{\alpha}_{\ a}$ and $\chi_{\alpha}=0$. 
These covariant gauge-fixing conditions are given by 
the equations of motion for the auxiliary fields ${\cal B}_{1\alpha}^{\ a}$ and 
${\cal B}^{\alpha}_{2}$. 
Using these gauge-fixing conditions, the zweibein field 
and the gravitino field vanish from the action,
so that the action has only $X^{\mu}$ and $\psi^{\mu}_{A}$.
After gauge fixing, we use the world-sheet light-cone coordinates 
$\sigma^{\pm}=\tau\pm\sigma$, so that 
the components of the world-sheet metric and 
the world-sheet totally antisymmetric tensor become
$\eta^{+-}=\eta^{-+}=-2$ and $\epsilon^{+-}=-\epsilon^{-+}=-2$.
Also, their partial derivatives are then 
$\partial_{\pm}=\frac{1}{2}(\partial_{\tau}\pm\partial_{\sigma})$.
Moreover, we use the spacetime light-cone coordinates
$X^{\pm}=\frac{1}{\sqrt{2}}(X^{0}\pm X^{1})$ 
and the spacetime light-cone components of Majorana fermion 
$\psi^{\pm}_{A}=\frac{1}{\sqrt{2}}(\psi^{0}_{A}\pm \psi^{1}_{A})$.

The condition that superstring theory be Weyl-invariant in quantization on
the world-sheet requires that the renormalization group $\beta$ functions
must vanish at all loop orders; 
these necessary conditions correspond to the field equations,
which resembles Einstein's equation, the antisymmetric
tensor generalization of Maxwell's equation and so on\cite{Polchinski}.
As the background fields which satisfy these field equations, 
we use the following $\it pp$-wave metric and antisymmetric
tensor field, whose flux is a constant:
\begin{align}
ds^2&=-\mu^2(X^2+Y^2)dX^{+}dX^{+}-2dX^{+}dX^{-}+ dXdX+dYdY
+dX^{k}dX^{k}\,\,, \\
B&=-\mu YdX^{+}\wedge dX+\mu XdX^{+}\wedge dY\,\,,
\end{align}
where we define $X^{\mu=2}=X$ and $X^{\mu=3}=Y$, 
and the index $k$ runs over $4\,,\,5\,,\,\cdots\,,\,D-1$, 
that is to say, the components of $G_{\mu\nu}$ and $B_{\mu\nu}$ are
\begin{align}
G_{++}&=-\mu^2(X^2+Y^2)\,,
\,\,\,\,\,
G_{+-}=G_{-+}=-1 \,, \\
G_{ij}&=\delta_{ij}\,,\,\,\,\,\,\, i,j=2,3,\cdots, D-1\,, \\
B_{+2}&=-B_{2+}=-\mu\,Y
\,,\,\,\,\,\,\,\,\,
B_{+3}=-B_{3+}=\mu\,X\,,
\end{align}
with all others vanishing. 
In this NS-NS {\it pp}-wave background, 
the generalized Riemann tensor becomes
${\cal R}_{\mu\rho\nu\sigma}=0$ because of the existence of the NS-NS flux, 
however this spacetime is highly curved at the standpoint of 
the ordinary Riemann tensor $R_{\mu\rho\nu\sigma}$. 
Finally, we introduce the complex coordinates $Z=X+iY$, $Z^{*}=X-iY$
\footnote{In our previous paper\cite{Chi-Ya}, we used the notation 
of $\bar{Z}$ for the complex conjugate of $Z$, 
however in this paper we use the notation of bar for the Dirac conjugate 
of the fermionic field. 
Therefore we use the notation of $Z^{*}$ for the complex conjugate of 
$Z$ in this paper.},
and the spacetime complex components of Majorana fermion 
$\psi^{Z}_{A}=\psi^{2}_{A}+i\psi^{3}_{A}$, $\psi^{Z*}_{A}
=\psi^{2}_{A}-i\psi^{3}_{A}$. 
Using the complex coordinates the {\it pp}-wave metric become 
$G_{++}=-\mu^{2}Z^{*}Z,\ G_{ZZ^{*}}=G_{Z^{*}Z}=\frac{1}{2}$ 
and the NS-NS field become 
$B_{+Z}=-B_{Z+}=-\frac{i}{2}\mu Z^{*},\ B_{+Z^{*}}=-B_{Z^{*}+}=\frac{i}{2}\mu Z$. 
Moreover we introduce the world-sheet covariant derivatives 
$D_{\pm}=\partial_{\pm}\pm i\mu\partial_{\pm}X^{+}$.
Then the action $S_{\rm M}$ becomes the following form:
\begin{align}
S_{\rm M}=\frac{1}{2\pi\alpha'}\int d\tau d\sigma
\Bigl[
&-2\partial_{+}X^{+}\partial_{-}X^{-}-2\partial_{-}X^{+}\partial_{+}X^{-}
+(D_{+}Z)^{*}D_{-}Z
+(D_{-}Z)^{*}D_{+}Z+2\partial_{+}X^{k}\partial_{-}X^{k}
\nonumber\\
&-i\psi^{+}_{+}\partial_{-}\psi^{-}_{+}-i\psi^{+}_{-}\partial_{+}\psi^{-}_{-}
-i\psi^{-}_{+}\partial_{-}\psi^{+}_{+}-i\psi^{-}_{-}\partial_{+}\psi^{+}_{-}
\nonumber\\
&+\frac{i}{2}\left\{\psi^{Z*}_{+}D_{-}\psi^{Z}_{+}
+\psi^{Z}_{+}(D_{-}\psi^{Z}_{+})^{*}+\psi^{Z*}_{-}D_{+}\psi^{Z}_{-}
+\psi^{Z}_{-}(D_{+}\psi^{Z}_{-})^{*}\right\}\nonumber\\
&+\mu\psi^{+}_{+}\left\{\psi^{Z*}_{+}D_{-}Z-\psi^{Z}_{+}(D_{-}Z)^{*}\right\}
-\mu\psi^{+}_{-}\left\{\psi^{Z*}_{-}D_{+}Z-\psi^{Z}_{-}(D_{+}Z)^{*}\right\}
\nonumber\\
&-i\mu^{2}Z^{*}Z(\psi^{+}_{+}\partial_{-}\psi^{+}_{+}
+\psi^{+}_{-}\partial_{+}\psi^{+}_{-})+i\psi^{k}_{+}\partial_{-}\psi^{k}_{+}
+i\psi^{k}_{-}\partial_{+}\psi^{k}_{-}
\Big]\,,
\label{action-X}
\end{align}
where we remove $(\psi^{+}_{\pm})^{2}$ from the action $S_{\rm M}$ 
because of $(\psi^{+}_{\pm})^{2}=0$ in both quantum theory and classical theory.
Here we note that the upper indices of $\psi^{\mu}_{A}$ 
always denote spacetime indices and lower indices of $\psi^{\mu}_{A}$ 
always denote world-sheet spinor indices. Do not confuse upper indices 
with lower indices.


\section{The equations of motion of $X^{\mu}$ and $\psi^{\mu}_{A}$ and 
their general solutions}
We obtain the equations of motion of $X^{\mu}$ and $\psi^{\mu}_{A}$ from 
the action (\ref{action-X})
\footnote{When we differentiate with the Grassmann number $\theta$, 
we always differentiate from the right-hand side:
\begin{align}
\frac{{\partial}(AB)}{{\partial}\theta}=A\frac{\partial B}{\partial\theta}
+(-1)^{|B|}\frac{\partial A}{\partial\theta}B\nonumber
\end{align}
where $|B|$ is the statistical factor of $B$}.
These equations are obviously related to the Heisenberg equations 
of motion with respect to quantization. 
The equations of $X^{\mu}$ and $\psi^{\mu}_{A}$ intricately interact 
with each other in the simple variation of the action. 
However we can remove fermionic fields $\psi^{\mu}_{A}$ 
from the equations of $X^{\mu}$, using the equations of $\psi^{\mu}_{A}$. 
Therefore the equations of 
$X^{\mu}$ become the same as the equations in our previous paper\cite{Chi-Ya}. 
It is no exaggeration to say that this enables us easily to quantize the case of 
the superstring in the NS-NS {\it pp}-wave background. 
Then we obtain the final equations of motion as follows.

$\bullet\,\,$ The equations of motion of $X^{+}$ and $X^{k}$ are
\begin{align}
\partial_{+}\partial_{-}X^{+}=0\,,
\,\,\,\,\,
\quad\partial_{+}\partial_{-}X^{k}=0\,.
\label{eom:X+k}
\end{align}

$\bullet\,\,$ The equations of motion of $Z$ and $Z^{*}$ are
\begin{align}
D_{+}D_{-}Z=0\,,
\,\,\,\,\,
\quad D^{*}_{+}D^{*}_{-}Z^{*}=0\,.
\label{eom:Z}
\end{align}

$\bullet\,\,$ The equation of motion of $X^{-}$ is
\begin{align}
\partial_{+}\partial_{-}X^{-}+\frac{i\mu}{4}
\left[
\partial_{+}\left(Z^{*}D_{-}Z-Z D^{*}_{-}Z^{*}\right)
-\partial_{-}\left(Z^{*}D_{+}Z-Z D^{*}_{+}Z^{*}\right)
\right]
=0\,.
\label{eom:X-}
\end{align}

$\bullet\,\,$ The equations of motion of 
$\psi^{+}_{\pm}$ and $\psi^{k}_{\pm}$ are
\begin{align}
\partial_{-}\psi^{+}_{+}=0,\,\,\,\,\,\partial_{+}\psi^{+}_{-}=0,\,\,\,\,\,
\partial_{-}\psi^{k}_{+}=0,\,\,\,\,\,\partial_{+}\psi^{k}_{-}=0\,.
\end{align}

$\bullet\,\,$ The equations of motion of $\psi^{Z}_{\pm}$ 
and $\psi^{Z*}_{\pm}$ are
\begin{align}
D_{-}(\psi^{Z}_{+}+i\mu\psi^{+}_{+}Z)&=0,\,\,\,\,\,
D_{+}(\psi^{Z}_{-}-i\mu\psi^{+}_{-}Z)=0,
\label{Psi-Z+-}
\\
D^{*}_{-}(\psi^{Z*}_{+}-i\mu\psi^{+}_{+}Z^{*})&=0,\,\,\,\,\,
D^{*}_{+}(\psi^{Z*}_{-}+i\mu\psi^{+}_{-}Z^{*})=0\,.
\end{align}

$\bullet\,\,$ The equations of motion of $\psi^{-}_{\pm}$ are
\begin{align}
\partial_{-}\psi^{-}_{+}+\frac{i}{2}
\mu\left[\psi^{Z*}_{+}D_{-}Z-\psi^{Z}_{+}(D_{-}Z)^{*}\right]
+\frac{1}{2}\mu^{2}\partial_{-}(Z^{*}Z)\psi^{+}_{+}&=0,\\
\partial_{+}\psi^{-}_{-}-\frac{i}{2}\mu\left[\psi^{Z*}_{-}D_{+}Z
-\psi^{Z}_{-}(D_{+}Z)^{*}\right]+\frac{1}{2}\mu^{2}\partial_{+}(Z^{*}Z)
\psi^{+}_{-}&=0\,.
\end{align}

Firstly we can solve the equations of bosonic fields, using the same method of 
the previous paper\cite{Chi-Ya}. 
The general solutions of bosonic fields are as follows:
\begin{align}
X^{+}(\tau,\sigma)&=X^{+}_{\rm L}(\sigma^{+})+X^{+}_{\rm R}(\sigma^{-}),
\label{1}\\
X^{k}(\tau,\sigma)&=X^{k}_{\rm L}(\sigma^{+})+X^{k}_{\rm R}(\sigma^{-}),\\
Z(\tau,\sigma)&=e^{-i\mu\tilde{X}^{+}}\left[f(\sigma^{+})+g(\sigma^{-})\right],
\\
Z^{*}(\tau,\sigma)&=e^{i\mu\tilde{X}^{+}}\left[f^{*}(\sigma^{+})
+g^{*}(\sigma^{-})\right],\\
X^{-}(\tau,\sigma)&=X^{-}_{\rm L}(\sigma^{+})+X^{-}_{\rm R}(\sigma^{-})
+\frac{i}{2}
\left[f(\sigma^{+})g^{*}(\sigma^{-})-f^{*}(\sigma^{+})g(\sigma^{-})\right]\,,
\label{X}
\end{align}
where L and R indicate the left-moving and right-moving parts, respectively.
Here we define 
\begin{align}
\tilde{X}^{+}=X^{+}_{\rm L}-X^{+}_{\rm R}.
\end{align}
We note that $\tilde{X}^{+}$ is not the periodic function, 
so that the arbitrary functions $f(\sigma^{+})$ and $g(\sigma^{-})$
satisfy the twisted boundary conditions.

Secondly we solve the equations of fermionic fields. 
The easiest equations are $\psi^{+}_{\pm}$ and $\psi^{k}_{\pm}$. 
Thus the solutions are
\begin{align}
\psi^{+}_{+}(\tau,\sigma)&=\psi^{+}_{+}(\sigma^{+}),\,\,\,\,
\psi^{+}_{-}(\tau,\sigma)=\psi^{+}_{-}(\sigma^{-}),
\label{2}\\
\psi^{k}_{+}(\tau,\sigma)&=\psi^{k}_{+}(\sigma^{+}),\,\,\,\,
\psi^{k}_{-}(\tau,\sigma)=\psi^{k}_{-}(\sigma^{-}).
\label{Psi+k}
\end{align}
Needless to say, these fields are free fields. 
In the next place we solve the equations of $\psi^{Z}_{\pm}$ 
and $\psi^{Z*}_{\pm}$. 
Because we can obtain the solutions of $\psi^{Z*}_{\pm}$ 
from Hermitian conjugate of $\psi^{Z}_{\pm}$,  
it is enough to solve the equations of $\psi^{Z}_{\pm}$.
Here we define $\Psi^{Z}_{\pm}=\psi^{Z}_{\pm}\pm i\mu\psi^{+}_{\pm}Z$. 
The differential equations of (\ref{Psi-Z+-}) become easier:
\begin{align}
(\partial_{\mp}\mp i\mu\partial_{\mp}X^{+})\Psi^{Z}_{\pm}=0.
\end{align}
The general solutions are
\begin{align}
\Psi^{Z}_{\pm}=e^{-i\mu\tilde{X}^{+}}\lambda_{\pm}(\sigma^{\pm})\,,
\end{align}
where $\lambda_{+}$ and $\lambda_{-}$ are
arbitrary Grassmann function of $\sigma^{+}$ 
and arbitrary Grassmann function of $\sigma^{-}$ respectively.
Here we may find the solutions which have only 
$e^{-i\mu X^{+}_{\rm L}}$ or $e^{-i\mu X^{+}_{\rm R}}$; 
however, these solutions are difficult to quantize, 
and we had better choose the solutions which have the same the factor as $Z$ 
from the standpoint of supersymmetry.
Therefore we obtain the general solutions of $\psi^{Z}_{\pm}$ 
and $\psi^{Z*}_{\pm}$:
\begin{align}
\psi^{Z}_{\pm}(\tau,\sigma)&=e^{-i\mu\tilde{X}^{+}}\lambda_{\pm}(\sigma^{\pm})
\mp i\mu\psi^{+}_{\pm}(\sigma^{\pm})Z(\tau,\sigma),\label{general_sol:psi^Z}\\
\psi^{Z*}_{\pm}(\tau,\sigma)&=e^{i\mu\tilde{X}^{+}}
\lambda^{*}_{\pm}(\sigma^{\pm})\pm i\mu\psi^{+}_{\pm}(\sigma^{\pm})
Z^{*}(\tau,\sigma)\,,
\label{general_sol:psi^Z*}
\end{align}
where $\psi^{Z*}_{\pm}$ is obtained by taking the Hermitian conjugate of 
$\psi^{Z}_{\pm}$. Moreover substituting the general solution of $Z$ and $Z^{*}$, 
we can represent $\psi^{Z}_{\pm}$ and $\psi^{Z*}_{\pm}$ 
by using the twisted fields $f$ and $g$:
\begin{align}
\psi^{Z}_{+}(\tau,\sigma)&=e^{-i\mu\tilde{X}^{+}}\left[\lambda_{+}(\sigma^{+})-i\mu\psi^{+}_{+}(\sigma^{+})\left\{f(\sigma^{+})+g(\sigma^{-})\right\}\right],\\
\psi^{Z}_{-}(\tau,\sigma)&=e^{-i\mu\tilde{X}^{+}}\left[\lambda_{-}(\sigma^{-})
+i\mu\psi^{+}_{-}(\sigma^{-})
\left\{f(\sigma^{+})+g(\sigma^{-})\right\}\right],\\
\psi^{Z*}_{+}(\tau,\sigma)&=e^{i\mu\tilde{X}^{+}}
\left[\lambda^{*}_{+}(\sigma^{+})+i\mu\psi^{+}_{+}(\sigma^{+})
\left\{f^{*}(\sigma^{+})+g^{*}(\sigma^{-})\right\}\right],\\
\psi^{Z*}_{-}(\tau,\sigma)&=e^{i\mu\tilde{X}^{+}}\left[\lambda^{*}_{-}(\sigma^{-})-i\mu\psi^{+}_{-}(\sigma^{-})\left\{f^{*}(\sigma^{+})+g^{*}(\sigma^{-})\right\}\right].
\end{align}
We can confirm the existence of the supersymmetry between $Z$ and 
$\psi^{Z}_{\pm}$ from direct calculation. 

Finally we solve the equations of $\psi^{-}_{\pm}$. 
As a matter of fact, using the equations of motion of 
$\psi^{+}_{\pm},\ \psi^{Z}_{\pm}$, the equations of motion of $\psi^{-}_{\pm}$ 
become easier forms:
\begin{align}
\partial_{-}\psi^{-}_{+}&=\partial_{-}\left[-\frac{i}{2}(Z\psi^{Z*}_{+}
-Z^{*}\psi^{Z}_{+})-\mu^{2}\psi^{+}_{+}Z^{*}Z\right],\\
\partial_{+}\psi^{-}_{-}&=\partial_{+}\left[\frac{i}{2}(Z\psi^{Z*}_{-}
-Z^{*}\psi^{Z}_{-})-\mu^{2}\psi^{+}_{-}Z^{*}Z\right].
\end{align}
Therefore we can solve these equations, removing the differential operator from
the left-hand side. 
The general solutions of $\psi^{-}_{\pm}$ are 
\begin{align}
\psi^{-}_{+}&=\psi^{\ -}_{0+}-\frac{i}{2}(Z\psi^{Z*}_{+}-Z^{*}\psi^{Z}_{+})
-\mu^{2}\psi^{+}_{+}Z^{*}Z\,,\label{general sol:psi-+}\\
\psi^{-}_{-}&=\psi^{\ -}_{0-}+\frac{i}{2}(Z\psi^{Z*}_{-}-Z^{*}\psi^{Z}_{-})
-\mu^{2}\psi^{+}_{-}Z^{*}Z\label{general sol:psi--}\,,
\end{align}
where $\psi^{\ -}_{0+}$ and $\psi^{\ -}_{0-}$ are an arbitrary Grassmann function
of $\sigma^{+}$ and an arbitrary Grassmann function of $\sigma^{-}$ respectively.
Moreover substituting the general solutions of 
$Z,\ Z^{*}, \psi^{Z}_{\pm},\ \psi^{Z*}_{\pm}$, 
we can represent $\psi^{-}_{\pm}$ by using the twisted fields 
$f,\ g,\ \lambda_{\pm}$ 
as follows:
\begin{align}
\psi^{-}_{+}(\tau,\sigma)&=\psi^{\ -}_{0+}(\sigma^{+})
-\frac{i}{2}\mu\left[f(\sigma^{+})+g(\sigma^{-})\right]
\lambda^{*}_{+}(\sigma^{+})
+\frac{i}{2}\mu\left[f^{*}(\sigma^{+})+g^{*}(\sigma^{-})\right]
\lambda_{+}(\sigma^{+}),
\label{4}\\
\psi^{-}_{-}(\tau,\sigma)&=\psi^{\ -}_{0-}(\sigma^{-})
+\frac{i}{2}\mu\left[f(\sigma^{+})+g(\sigma^{-})\right]
\lambda^{*}_{-}(\sigma^{-})
-\frac{i}{2}\mu\left[f^{*}(\sigma^{+})+g^{*}(\sigma^{-})\right]
\lambda_{-}(\sigma^{-})\,.
\label{Psi-}
\end{align}

We can solve all the equations of motion of the superstring 
in the NS-NS {\it pp}-wave background generally and we represent 
all the general solutions by using the useful fields
in spite of existence of interactions. 
After this section, we quantize these general solutions. 
Before we quantize these solutions, we have to consider the method 
to quantize superstrings in the background fields. 
Superstrings in the background fields have some constraints. 
In the following section, we consider the constraints and we construct 
the formula of canonical (anti)commutation relations by using 
the Dirac brackets.


\section{Constraints from Majorana fermion and Dirac bracket}
We consider the second-class constraints from RNS superstrings 
in the background fields $G_{\mu\nu}$ and $B_{\mu\nu}$. 
The canonical momenta associated with string coordinates $X^{\mu}$ 
and Majorana fermions $\psi^{\mu}_{A}$  are defined as 
\begin{align}
P_{\mu}&=\frac{\partial S_{\rm M}}{\partial(\partial_{\tau}X^{\mu})}
=\frac{1}{2\pi\alpha'}\left[G_{\mu\nu}\partial_{\tau}X^{\nu}
-B_{\mu\nu}\partial_{\sigma}X^{\nu}+\frac{i}{2}
G_{\rho\omega}\psi^{\dagger\rho}
(\Gamma^{\omega}_{\mu\nu}+\frac{\gamma_{3}}{2}
H^{\omega}_{\mu\nu})\psi^{\nu}
\right]\,,\\
\pi_{\mu A}&=\frac{\partial S_{\rm M}}{\partial(\partial_{\tau}\psi^{\mu}_{A})}
=\frac{i}{4\pi\alpha'}G_{\mu\nu}\psi^{\nu}_{A}\,,
\end{align}
where $\partial_{\tau}=\frac{\partial}{\partial \tau}$ and
$\partial_{\sigma}=\frac{\partial}{\partial \sigma}$.
In the case of the fermions, we always differentiate the action 
with the Grassmann number from right-hand side.
Although the momenta $P_{\mu}$ contain $\partial_{\tau}X^{\mu}$ certainly,
the momenta $\pi_{\mu A}$ do not contain $\partial_{\tau}\psi^{\mu}_{A}$.
Therefore, they give rise to the second-class constraints $\phi_{\mu A}=\pi_{\mu A}-\frac{i}{4\pi\alpha'}G_{\mu\nu}\psi^{\nu}_{A}\approx 0$. 
Here we must note that $G_{\mu\nu}$ is a function of string coordinates 
$X^{\mu}$. 
For this reason, Poisson brackets between the canonical momenta 
$P_{\mu}$ and the constraints $\phi_{\mu A}$ are not zero,
therefore we also must consider the Dirac brackets about $P_{\mu}$. 
The Poisson brackets between $X^{\mu}$ and $\phi_{\mu A}$ vanish, 
so that the Dirac brackets which contain $X^{\mu}$ 
become the Poisson brackets.
The other Dirac brackets become as follows:
\footnote{
In this paper, we leave out the description of $\tau$ in the right-hand side 
about the Dirac brackets and the canonical (anti)commutation relations.
}
\begin{align}
\{\psi^{\mu}_{A}(\tau,\sigma),\psi^{\nu}_{B}(\tau,\sigma')\}_{\rm D}
&=
\begin{cases}
-i\cdot2\pi\alpha'G^{\mu\nu}\delta_{AB}\delta(\sigma-\sigma') 
\qquad \text{: R sector}\,,\\
-i\cdot2\pi\alpha'G^{\mu\nu}\delta_{AB}\tilde{\delta}(\sigma-\sigma') 
\qquad \text{: NS sector}\,,
\end{cases}\\
\{\psi^{\mu}_{A}(\tau,\sigma),\pi_{\nu B}(\tau,\sigma')\}_{\rm D}
&=
\begin{cases}
\frac{1}{2}\delta^{\mu}_{\nu}\delta_{AB}\delta(\sigma-\sigma') 
\ \ \ \ \ \ \ \ \ \ \ \ \qquad \text{: R sector}\,,\\
\frac{1}{2}\delta^{\mu}_{\nu}\delta_{AB}\tilde{\delta}(\sigma-\sigma') 
\ \ \ \ \ \ \ \ \ \ \ \ \qquad \text{: NS sector}\,,
\end{cases}\\
\{\pi_{\mu A}(\tau,\sigma),\pi_{\nu B}(\tau,\sigma')\}_{\rm D}
&=
\begin{cases}
\frac{i}{8\pi\alpha'}G_{\mu\nu}\delta_{AB}\delta(\sigma-\sigma') 
\ \ \ \ \ \ \qquad \text{: R sector}\,,\\
\frac{i}{8\pi\alpha'}G_{\mu\nu}\delta_{AB}\tilde{\delta}(\sigma-\sigma') 
\ \ \ \ \ \ \qquad \text{: NS sector}\,,
\end{cases}\\
[P_{\mu}(\tau,\sigma),P_{\nu}(\tau,\sigma')]_{\rm D}
&=-\frac{i}{8\pi\alpha'}\partial_{\mu}G_{\rho\lambda}
\partial_{\nu}G_{\sigma\omega}G^{\rho\sigma}\psi^{\lambda}_{A}(\sigma')
\psi^{\omega}_{A}(\sigma')
\delta(\sigma-\sigma')\,,\\
[P_{\mu}(\tau,\sigma),\psi^{\nu}_{A}(\tau,\sigma')]_{\rm D}
&=\frac{1}{2}\partial_{\mu}G_{\rho\sigma}G^{\rho\nu}\psi^{\sigma}_{A}(\sigma')
\delta(\sigma-\sigma')\,,\\
[P_{\mu}(\tau,\sigma),\pi_{\nu A}(\tau,\sigma')]_{\rm D}
&=-\frac{i}{8\pi\alpha'}\partial_{\mu}G_{\nu\rho}\psi^{\rho}_{A}(\sigma')
\delta(\sigma-\sigma')\,.
\end{align}
Here $\delta(\sigma)$ denotes the ordinary periodic delta-function and 
$\tilde{\delta}(\sigma)$ denotes the anti-periodic delta-function.
We must consider the difference between the Ramond (R) fermions 
and the Neveu-Schwarz (NS) fermions. 
We must especially take notice of the NS fermions which 
have the anti-periodicity.
In the case of the R fermions we can identify the world-sheet coordinates 
$\sigma$ and $\sigma'$ using the property of the periodic delta-function, 
however in the case of the NS fermions we cannot do it simply 
because of the anti-periodicity. 
The relation between the NS fermion and the delta-functions is 
\begin{align}
\psi^{\mu}(\sigma')\delta(\sigma-\sigma')=
\psi^{\mu}(\sigma)\tilde{\delta}(\sigma-\sigma')\,.\label{Rel-NS-delta}
\end{align}
We can prove this relation (\ref{Rel-NS-delta}) by using the Fourier expansions. 
Substituting the condition 
$\pi_{\mu A}=\frac{i}{4\pi\alpha'}G_{\mu\nu}\psi^{\nu}_{A}$ to
the Dirac brackets 
$\{\psi^{\mu}_{A}(\tau,\sigma),\pi_{\nu B}(\tau,\sigma')\}_{\rm D}$ and 
$\{\pi_{\mu A}(\tau,\sigma),\pi_{\nu B}(\tau,\sigma')\}_{\rm D}$,
these correspond to the Dirac bracket 
$\{\psi^{\mu}_{A}(\tau,\sigma),\psi^{\nu}_{B}(\tau,\sigma')\}_{\rm D}$.
Moreover using the Poisson bracket between $P_{\mu}$ and $G_{\mu\nu}$ 
and the Leibniz rule, the Dirac bracket 
$[P_{\mu}(\tau,\sigma),\pi_{\nu A}(\tau,\sigma')]_{\rm D}$ corresponds 
to the Dirac bracket $[P_{\mu}(\tau,\sigma),\psi^{\nu}_{A}(\tau,\sigma')]_{\rm D}$. 
They mean that we have only to consider the Dirac brackets between 
$X^{\mu},\ P_{\mu},\ \psi^{\mu}_{A}$. 
Therefore we do not have to consider the Dirac bracket about $\pi_{\mu A}$. 

We quantize this constraint system, using the ordinary procedure 
which we replace $i$ times Dirac brackets
with the canonical (anti)commutation relations.
Let us treat the case of the {\it pp}-wave background. 
The canonical (anti)commutation relations are as follows:

$\bullet\,\,$ The canonical commutation relations between the bosonic fields 
and their momenta are
\begin{align}
\left[X^{\mu}(\tau,\sigma),P_{\nu}(\tau,\sigma')\right]
\label{canonical-com-rel:XP}
&=i\delta^{\mu}_{\ \nu}\delta(\sigma-\sigma')\,,\\
\left[X^{\mu}(\tau,\sigma),X^{\nu}(\tau,\sigma')\right]=0\,, \qquad
&\left[P_{\mu}(\tau,\sigma),P_{\nu}(\tau,\sigma')\right]=0\,
\label{canonical-com-rel:XXPP}.
\end{align}

$\bullet\,\,$ The canonical anticommutation relations between 
the fermionic fields are
\begin{align}
\{\psi^{-}_{A}(\tau,\sigma),\psi^{-}_{B}(\tau,\sigma')\}
&=
\begin{cases}
2\pi\alpha'\mu^{2}Z^{*}Z\delta_{AB}\delta(\sigma-\sigma') 
\qquad \text{: R sector}\,,\\
2\pi\alpha'\mu^{2}Z^{*}Z\delta_{AB}\tilde{\delta}(\sigma-\sigma') 
\qquad \text{: NS sector}\,,
\end{cases}
\label{canonical-anti-com-rel:psi-psi-}\\
\{\psi^{+}_{A}(\tau,\sigma),\psi^{-}_{B}(\tau,\sigma')\}
&=
\begin{cases}
-2\pi\alpha'\delta_{AB}\delta(\sigma-\sigma') 
\ \ \ \ \ \ \ \qquad \text{: R sector}\,,\\
-2\pi\alpha'\delta_{AB}\tilde{\delta}(\sigma-\sigma') 
\ \ \ \ \ \ \ \qquad \text{: NS sector}\,,
\end{cases}\label{canonical-anti-com-rel:psi+psi-}\\
\{\psi^{Z}_{A}(\tau,\sigma),\psi^{Z*}_{B}(\tau,\sigma')\}
&=
\begin{cases}
4\pi\alpha'\delta_{AB}\delta(\sigma-\sigma') 
\ \ \ \ \ \ \ \ \ \ \qquad \text{: R sector}\,,\\
4\pi\alpha'\delta_{AB}\tilde{\delta}(\sigma-\sigma') 
\ \ \ \ \ \ \ \ \ \ \qquad \text{: NS sector}\,,
\end{cases}\label{canonical-anti-com-rel:psiZpsiZ*}\\
\{\psi^{k}_{A}(\tau,\sigma),\psi^{l}_{B}(\tau,\sigma')\}
&=
\begin{cases}
2\pi\alpha'\delta^{kl}\delta_{AB}\delta(\sigma-\sigma') 
\ \ \ \ \ \ \qquad \text{: R sector}\,,\\
2\pi\alpha'\delta^{kl}\delta_{AB}\delta(\sigma-\sigma') 
\ \ \ \ \ \ \qquad \text{: NS sector}\,.
\end{cases}\label{canonical-anti-com-rel:psi-k}
\end{align}

$\bullet\,\,$ The canonical commutation relations between the bosonic fields 
and  the fermionic fields are
\begin{align}
[X^{\mu}(\tau,\sigma),\psi^{\nu}_{A}(\tau,\sigma')]&=0,
\label{canonical-com-rel:X-psi}\\
[P_{Z}(\tau,\sigma),\psi^{-}_{A}(\tau,\sigma')]
&=\frac{i}{2}\mu^{2}Z^{*}\psi^{+}_{A}(\sigma')\delta(\sigma-\sigma'),
\label{canonical-com-rel:PZpsi-}\\
[P_{Z^{*}}(\tau,\sigma),\psi^{-}_{A}(\tau,\sigma')]
&=\frac{i}{2}\mu^{2}Z\psi^{+}_{A}(\sigma')\delta(\sigma-\sigma')\,,
\label{canonical-com-rel:PZ*psi-}
\end{align}
with all other commutations vanishing.
In the case of the NS fermion, the world-sheet coordinate of fermion must be 
$\sigma'$
because of the anti-periodicity. 
The characteristic point is that the (anti)commutation relations between 
$\psi^{-}_{A},\ P_{Z}$ and $P_{Z^{*}}$ do not vanish. 
We note that fortunately the commutation relations between the momenta vanish 
because of the inverse of metric $G^{++}=0$ in the {\it pp}-wave background. 
Therefore we can quantize the bosonic fields using the same method 
of the previous paper\cite{Chi-Ya}. 
Of course when $\mu=0$, we obtain ordinary canonical (anti)commutation 
relations of superstrings in the flat spacetime.


\section{Free-mode representation}
In this section we construct the free-mode representations in which 
the general operator solutions satisfy all the canonical (anti)commutation 
relations.
The detailed proof of the free-mode representations 
in the canonically covariant quantization is given in the next section.
The free-mode representations of bosonic fields without $X^{-}$ are 
the same as the previous paper\cite{Chi-Ya}. 
Because $X^{-}$ interact with $\psi^{Z}_{A}$ and $\psi^{Z*}_{A}$, 
we have to newly construct the free-mode representation of $X^{-}$ 
from the commutation relations between 
$X^{-}$ and $\psi^{Z}_{A}$, $\psi^{Z*}_{A}$. 
The free-mode representations of bosonic fields $X^{+},\ X^{k},\ f,\ g$ 
are as follows:
\begin{align}
X^{+}&=x^{+}+\frac{\alpha'}{2}\,p^{+}(\sigma^{+}+\sigma^{-})
+i\sqrt{\frac{\alpha'}{2}}\sum_{n\neq 0}\frac{1}{n}
\left[
\tilde{\alpha}^{+}_{n}e^{-in\sigma^{+}}
+\alpha^{+}_{n}e^{-in\sigma^{-}}
\right]\,,\\
X^{k}&=x^{k}+\frac{\alpha'}{2}\,p^{k}(\sigma^{+}+\sigma^{-})
+i\sqrt{\frac{\alpha'}{2}}\sum_{n\neq 0}\frac{1}{n}
\left[
\tilde{\alpha}^{k}_{n}e^{-in\sigma^{+}}
+\alpha^{k}_{n}e^{-in\sigma^{-}}
\right]\,,\\
f(\sigma^{+})&=\sqrt{\alpha'}
\sum_{n\in{\mathbb Z}}\frac{A_{n}}{\sqrt{|n-\hat{\mu}|}}
e^{-i(n-\hat{\mu})\sigma^{+}},\\
g(\sigma^{-})&=\sqrt{\alpha'}
\sum_{n\in{\mathbb Z}}\frac{B_{n}}{\sqrt{|n+\hat{\mu}|}}
e^{-i(n+\hat{\mu})\sigma^{-}}.
\end{align}
where $n\neq 0$ which is put under $\Sigma$ means $n\in\mathbb{Z}(n\neq 0)$.
Here the dimensionless factor $\hat{\mu}$ is newly defined as
\begin{align}
\hat{\mu}=\mu\alpha'p^{+}\,.
\end{align}
We must note that $\hat{\mu}$ is not a constant, which depends 
on the momentum operator $p^{+}$, so that the commutation relation 
$[x^{-},\hat{\mu}]=-i\mu\alpha'$. 
By Hermitian conjugate of $f$ and $g$, we can obtain  $f^{*}$ and $g^{*}$,
whose modes are defined as  $A_{N}^{\dagger}$ and $B_{N}^{\dagger}$.
In the next place, we construct the free-mode representation of fermionic fields. 
First the free-mode representation of $\psi^{+}_{\pm}$ and $\psi^{k}_{\pm}$ are 
the same as ordinary free fields:
\begin{align}
\psi^{+}_{+}(\sigma^{+})&=\sqrt{\alpha'}\sum_{r\in{\mathbb Z}+\varepsilon}
\tilde{\psi}^{+}_{r}e^{-ir\sigma^{+}},\ \ \ \ 
\psi^{+}_{-}(\sigma^{-})=\sqrt{\alpha'}\sum_{r\in{\mathbb Z}+\varepsilon}
\psi^{+}_{r}e^{-ir\sigma^{-}},\\
\psi^{k}_{+}(\sigma^{+})&=\sqrt{\alpha'}\sum_{r\in{\mathbb Z}+\varepsilon}
\tilde{\psi}^{k}_{r}e^{-ir\sigma^{+}},\ \ \ \ 
\psi^{k}_{-}(\sigma^{-})=\sqrt{\alpha'}\sum_{r\in{\mathbb Z}+\varepsilon}
\psi^{k}_{r}e^{-ir\sigma^{-}}
\end{align}
where $\varepsilon=0$ in the R sector and $\varepsilon=\frac{1}{2}$ 
in the NS sector.
Second, we present the free-mode representations of 
$\psi^{Z}_{\pm}$ and $\psi^{Z*}_{\pm}$, which must satisfy the Ramond or 
Neveu-Schwarz boundary condition of the closed superstring theory. 
Because the factor $e^{-i\mu\tilde{X}^{+}(\tau,\sigma)}$ in $\psi^{Z}_{\pm}$ is 
transformed into $e^{-2\pi i\hat{\mu}}\cdot e^{-i\mu\tilde{X}^{+}(\tau,\sigma)}$ 
under the shift $\sigma\rightarrow\sigma+2\pi$, $\lambda_{\pm}$ 
in $\psi^{Z}_{\pm}$ 
are twisted fields. 
In other words, these fields satisfy the twisted boundary condition:
\begin{align}
\lambda_{+}(\sigma^{+}+2\pi)&=
\begin{cases}
+\,e^{2\pi i\hat{\mu}}\lambda_{+}(\sigma^{+})\ \qquad \text{: R sector}\,,\\
-\,e^{2\pi i\hat{\mu}}\lambda_{+}(\sigma^{+})\ \qquad \text{: NS sector}\,,\\
\end{cases}\label{boundary:lam+}\\
\lambda_{-}(\sigma^{-}-2\pi)&=
\begin{cases}
+\,e^{2\pi i\hat{\mu}}\lambda_{-}(\sigma^{-})\ \qquad \text{: R sector}\,,\\
-\,e^{2\pi i\hat{\mu}}\lambda_{-}(\sigma^{-})\ \qquad \text{: NS sector}\,.\\
\end{cases}\label{boundary:lam-}
\end{align}
Under the boundary conditions (\ref{boundary:lam+}), (\ref{boundary:lam-}), 
the free-mode representations of $\lambda_{\pm}$ are
\begin{align}
\lambda_{+}(\sigma^{+})&=\sqrt{2\alpha'}\sum_{r\in{\mathbb Z}
+\varepsilon}\tilde{\lambda}_{r}
e^{-i(r-\hat{\mu})\sigma^{+}},\label{free-rep:lam+}\\
\lambda_{-}(\sigma^{-})&=\sqrt{2\alpha'}\sum_{r\in{\mathbb Z}
+\varepsilon}\lambda_{r}e^{-i(r+\hat{\mu})\sigma^{-}}\label{free-rep:lam-}.
\end{align}
Thus we can obtain the free-mode representations of $\lambda^{*}_{\pm}$ 
by taking Hermitian conjugate of $\lambda_{\pm}$:
\begin{align}
\lambda^{*}_{+}(\sigma^{+})&=\sqrt{2\alpha'}\sum_{r\in{\mathbb Z}
+\varepsilon}\tilde{\lambda}^{\dagger}_{r}
e^{i(r-\hat{\mu})\sigma^{+}},\\
\lambda^{*}_{-}(\sigma^{-})&=\sqrt{2\alpha'}\sum_{r\in{\mathbb Z}
+\varepsilon}\lambda^{\dagger}_{r}
e^{i(r+\hat{\mu})\sigma^{-}}.
\end{align}
Third we present the free-mode representation of $\psi^{-}_{A}$. 
We have already known the free-mode representation of $f,\ g, \lambda_{\pm}$ 
and their Hermitan conjugate in $\psi^{-}_{A}$. 
Imposing the canonical (anti)commutation relations on 
$\psi^{-}_{A}$, $\psi_{0\pm}^{\ -}$ must be absolutely free field, 
so that $\psi^{\ -}_{0\pm}$ does not contain other fields. 
Therefore the free-mode representations of $\psi_{0\pm}^{\ -}$ are
the same as ordinary free fields:
\begin{align}
\psi_{0+}^{\ -}=\sqrt{\alpha'}\sum_{r\in{\mathbb Z}+\varepsilon}
\tilde{\psi}^{-}_{r}e^{-ir\sigma^{+}},\\
\psi_{0-}^{\ -}=\sqrt{\alpha'}\sum_{r\in{\mathbb Z}+\varepsilon}
\psi^{-}_{r}e^{-ir\sigma^{-}}.
\end{align}

Finally we explain the construction of the free-mode representation 
of $X^{-}$ in detail.
The canonical commutation relation between $X^{-}$ and $\psi^{Z}_{\pm}$ 
must be zero, therefore the commutation relations between $X^{-}$ 
and $\lambda_{\pm}$ 
must be 
\begin{align}
[X^{-}(\tau,\sigma),\lambda_{\pm}({\sigma'}^{\pm})]
=i\mu[X^{-}(\tau,\sigma),\tilde{X}^{+}(\sigma')]\lambda_{\pm}({\sigma'}^{\pm}).
\label{Comrel:X-lam}
\end{align}
We divide $X^{-}_{\rm L}+X^{-}_{\rm R}$ into an almost free part, $X^{-}_{0}$
\footnote{
`` almost free '' means that only the zero-mode of $X^{-}_{0}$ dose not commute 
with the twisted fields, namely $f,\ g,$ and $\lambda_{\pm}$. 
} 
and a completely non-free part, $X^{-}_{1}$. 
Moreover we divide $X^{-}_{1}$ into the bosonic part $X^{-}_{\rm 1B}$ 
which is constructed from $f$ and $g$ and the fermionic part $X^{-}_{\rm 1F}$ 
which is constructed from $\lambda_{\pm}$. 
The free-mode representation of $X^{-}_{\rm 1B}$ is the same as that of 
our previous paper,
and we can obtain $X^{-}_{\rm 1F}$ using the same method of 
our previous paper\cite{Chi-Ya}. 
Although we did not explain in detail how to construct 
the free-mode representation of $X^{-}$ in the previous paper,  
here, we explain how to construct it. 
The mode expansion of $X^{-}_{\rm1F}$ is 
\begin{align}
X^{-}_{\rm 1F}=\alpha'p^{-}_{\rm F}\tau+i\sqrt{\frac{\alpha'}{2}}
\sum_{n\neq 0}\frac{1}{n}\left[\tilde{\alpha}^{-}_{{\rm F}n}e^{-in\sigma^{+}}
+\alpha^{-}_{{\rm F}n}e^{-in\sigma^{-}}\right]
\label{X-F}.
\end{align}
Here we note that $X^{-}_{\rm 1F}$ do not contain the zero mode $x^{-}$,
which is only contained in $X^{-}_{0}$.
The commutation relation $[X^{-}(\tau,\sigma),\tilde{X}^{+}(\tau,\sigma')]$ is
\begin{align}
[X^{-}(\tau,\sigma),\tilde{X}^{+}(\tau,\sigma')]
&=[X^{-}_{0}(\tau,\sigma),\tilde{X}^{+}(\tau,\sigma')]\nonumber\\
&=-i\alpha'\sigma'+\alpha'\sum_{n\neq 0}\frac{1}{n}e^{in(\sigma-\sigma')}
\end{align}
Calculating commutation relation (\ref{Comrel:X-lam}) and comparing 
the left-hand side with the right-hand side, 
we can obtain the commutation relation between modes:
\begin{align}
[p^{-}_{\rm F},\tilde{\lambda}_{r}]&=-\mu\tilde{\lambda}_{r},
\ \ \ \ [p^{-}_{\rm F},\lambda_{r}]=\mu\lambda_{r},\\
[\tilde{\alpha}^{-}_{{\rm F}n},\tilde{\lambda}_{r}]&=-\mu\sqrt{2\alpha'}
\tilde{\lambda}_{n+r},\ \ \ \ 
[\alpha^{-}_{{\rm F}n},\lambda_{r}]=\mu\sqrt{2\alpha'}\lambda_{n+r}.
\end{align}
Therefore we can represent the modes 
$p^{-}_{\rm F},\ \tilde{\alpha}^{-}_{{\rm F}n},\ \alpha^{-}_{{\rm F}n}$ 
by the modes $\tilde{\lambda}_{r}$ and $\lambda_{r}$, 
using the anticommutation relations between the modes $\lambda_{r}$, 
$\lambda^{\dagger}_{r}$ (Eq.(\ref{ComRel:lam-r})) which are written up later:
\begin{align}
p^{-}_{\rm F}&=\mu\sum_{r\in{\mathbb Z}+\varepsilon}
(:\tilde{\lambda}^{\dagger}_{r}\tilde{\lambda}_{r}:
-:\lambda_{r}^{\dagger}\lambda_{r}:)\,,
\label{p-F}\\
\tilde{\alpha}^{-}_{{\rm F}n}&=\mu\sqrt{2\alpha'}
\sum_{r\in{\mathbb Z}+\varepsilon}
\tilde{\lambda}^{\dagger}_{r}\tilde{\lambda}_{n+r}\,,\ \ \ \ 
\alpha^{-}_{{\rm F}n}=-\mu\sqrt{2\alpha'}
\sum_{r\in{\mathbb Z}+\varepsilon}
\lambda^{\dagger}_{r}\lambda_{n+r}\,,
\label{alpha-F}
\end{align}
where the notation $:\: :$ represents the normal ordered product,
whose definition is given in the final part of this section 
(Eqs.(\ref{Normal-order-A})-(\ref{Normal-order-lambda})). 
Substituting these modes (\ref{p-F}), (\ref{alpha-F}) into (\ref{X-F}), 
we can obtain the free-mode representation of $X^{-}_{\rm 1F}$. 
Thus we can obtain the free-mode representation of $X^{-}$ 
using $X^{-}_{0}$, $X^{-}_{\rm 1B}$, $X^{-}_{\rm 1F}$ and 
$X^{-}_{2}$, where $X^{-}_{2}=\frac{i\mu}{2}(fg^{*}-f^{*}g)$:
\begin{align}
X^{-}=X^{-}_{0}+X^{-}_{\rm 1B}+X^{-}_{\rm 1F}+X^{-}_{2}\,.
\label{X-}
\end{align}
The almost free part $X^{-}_{0}$ is
\begin{align}
X^{-}_{0}=x^{-}+\frac{1}{2}\alpha'p^{-}(\sigma^{+}+\sigma^{-})
+i\sqrt{\frac{\alpha'}{2}}\sum_{n\neq 0}\frac{1}{n}\left[\tilde{\alpha}^{-}_{n}
e^{-in\sigma^{+}}+\alpha^{-}_{n}e^{-in\sigma^{-}}\right].
\end{align}
$X^{-}_{\rm 1B}$ is the same as that of our previous paper\cite{Chi-Ya}:
\begin{align}
X^{-}_{\rm 1B}&=\mu\alpha'
\sum_{n\in{\mathbb Z}}
\Bigl[\,
{\rm sgn}(n-\hat{\mu}):A_{n}^{\dagger}A_{n}:
-{\rm sgn}(n+\hat{\mu}):B_{n}^{\dagger}B_{n}:
\,\Bigr]
\tau\nonumber\\
&\;\;\;\; -\frac{i\mu\alpha'}{2}
\sum_{m\neq n}
\frac{1}{m-n}
\biggl[\,
\frac{m+n-2\hat{\mu}}{\omega^{+}_{m}\omega^{+}_{n}}
A^{\dagger}_{m}A_{n}\,e^{i(m-n)\sigma^{+}}
-\frac{m+n+2\hat{\mu}}{\omega^{-}_{m}\omega^{-}_{n}}
B_{m}^{\dagger}B_{n}\,e^{i(m-n)\sigma^{-}}
\,\biggr]\,, \label{X-1B}
\end{align}
where we define $\omega^{\pm}_{n}=\sqrt{|n\mp\hat{\mu}|}$. 
In the summation with $m\neq n$ in Eq.(\ref{X-1B}), 
$m$ and $n$ run from $-\infty$ to $\infty$, excluding $m=n$. 
Note that the terms in this summation are not influenced 
by the normal ordered product. 
From the above calculation (\ref{X-F})-(\ref{alpha-F}), 
we can obtain the free mode representation 
of $X^{-}_{\rm 1F}$:
\begin{align}
X^{-}_{\rm 1F}=\mu\alpha'\sum_{r\in{\mathbb Z}+\varepsilon}
(:\tilde{\lambda}^{\dagger}_{r}
\tilde{\lambda}_{r}:-:\lambda^{\dagger}_{r}\lambda_{r}:)\tau
+i\mu\alpha'\sum_{\stackrel{\scriptstyle n\neq 0}{r\in{\mathbb Z}+\varepsilon}}
\frac{1}{n}\left[\tilde{\lambda}^{\dagger}_{r}
\tilde{\lambda}_{n+r}e^{-in\sigma^{+}}-\lambda^{\dagger}_{r}\lambda_{n+r}
e^{-in\sigma^{-}}\right].\label{X-1F}
\end{align}
We can also represent $X^{-}_{\rm 1B}$ and $X^{-}_{\rm 1F}$ 
by using the fields $f,\ g$ and $\lambda_{\pm}$:
\begin{align}
X^{-}_{\rm 1B}&=\frac{i\mu}{2}
\biggl[
\int d\sigma^{+}:\left(f^{*}\partial_{+}f-\partial_{+}f^{*}f\right):
-\int d\sigma^{-}:\left(g^{*}\partial_{-}g
-\partial_{-}g^{*}g\right):
\biggr]
-\mu J_{\rm B}\sigma\,,\\
X^{-}_{\rm 1F}&=\frac{\mu}{2}\left[\int d\sigma^{+}:\lambda^{*}_{+}\lambda_{+}:
-\int d\sigma^{-}:\lambda^{*}_{-}\lambda_{-}:\right]-\mu J_{\rm F}\sigma.
\end{align}
Here the integrals are indefinite integrals, 
and we choose the constants of integration to be zero. 
Moreover $J_{\rm B}$ and $J_{\rm F}$ are
\begin{align}
J_{\rm B}&=\frac{i}{4\pi}\biggl[
\int^{2\pi}_{0} d\sigma^{+}:\left(f^{*}\partial_{+}f-\partial_{+}f^{*}f\right):
+\int^{2\pi}_{0} d\sigma^{-}:\left(g^{*}\partial_{-}g-\partial_{-}g^{*}g\right)
\biggr]\,,\\
J_{\rm F}&=\frac{1}{4\pi}\left[\int^{2\pi}_{0}d\sigma^{+}:
\lambda^{*}_{+}\lambda_{+}:
+\int^{2\pi}_{0}d\sigma^{-}:\lambda^{*}_{-}\lambda_{-}:\right]\,.
\end{align}
In the free-mode representation, $J_{\rm B}$ and $J_{\rm F}$ are given by
\begin{align}
J_{\rm B}&=\alpha'\sum_{n\in{\mathbb Z}}
\left[\,
{\rm sgn}(n-\hat{\mu}):A^{\dagger}_{n}A_{n}:
+{\rm sgn}(n+\hat{\mu}):B^{\dagger}_{n}B_{n}:
\,\right]\,,\\
J_{\rm F}&=\alpha'\sum_{r\in{\mathbb Z}+\varepsilon}
\left[:\tilde{\lambda}^{\dagger}_{r}\tilde{\lambda}_{r}:
+:\lambda^{\dagger}_{r}\lambda_{r}:\right]\,.
\end{align}
Substituting the free modes of $f$ and $g$ into $X^{-}_{2}$, it becomes
\begin{align}
X^{-}_{2}=\frac{i\mu\alpha'}{2}
\sum_{m,n\in{\mathbb Z}}
\frac{1}{\omega^{+}_{m}\omega^{-}_{n}}
\Bigl[\,
A_{m} B_{n}^{\dagger}\,e^{-i(m-n-2\hat{\mu})\tau-i(m-n)\sigma}
-A_{m}^{\dagger} B_{n}\,e^{i(m-n-2\hat{\mu})\tau+i(m-n)\sigma}
\,\Bigr]\,.
\end{align}
Therefore we can represent all the general operator solutions 
by using only free modes.

We now explicitly present the (anti)commutation relations for all the modes, 
in order to demonstrate that they are perfectly free-modes:

$\bullet\,\,$ The nonvanishing commutation relations between
the modes of bosonic fields are
\begin{align}
\left[x^{+},p^{-}\right]=\left[x^{-},p^{+}\right]=-i\,, \quad
\left[\tilde{\alpha}^{+}_{m},\tilde{\alpha}^{-}_{n}\right]=
\left[\alpha^{+}_{m},\alpha^{-}_{n}\right]=-m\delta_{m+n}\,,
\label{mode+-}
\\
[x^{k},p^{l}]=i\delta^{kl}\,, \quad
[\tilde{\alpha}^{k}_{m},\tilde{\alpha}^{l}_{n}]=
[\alpha^{k}_{m},\alpha^{l}_{n}]=m\delta^{kl}\delta_{m+n}\,,
\\
[A_{m},A^{\dagger}_{n}]={\rm sgn}(m-\hat{\mu})
\delta_{m,n}\,,\ \ \ \ \ \ \ \ 
[B_{m},B^{\dagger}_{n}]={\rm sgn}(m+\hat{\mu})
\delta_{m,n}
\,.
\end{align}

$\bullet\,\,$ The nonvanishing anticommutation relations between the modes of 
$\psi^{+}_{\pm}$ and $\psi^{\ -}_{0\pm}$ are
\begin{align}
\{\tilde{\psi}^{+}_{r},\tilde{\psi}^{-}_{s}\}
=\{\psi^{+}_{r},\psi^{-}_{s}\}=-\delta_{r+s}.
\end{align}

$\bullet\,\,$ The nonvanishing anticommutation relations between the modes of 
$\lambda_{\pm}$ and $\lambda^{*}_{\pm}$ are
\begin{align}
\{\tilde{\lambda}_{r},\tilde{\lambda}^{\dagger}_{s}\}
=\{\lambda_{r},\lambda^{\dagger}_{s}\}=\delta_{r-s}\label{ComRel:lam-r}.
\end{align}

$\bullet\,\,$ The nonvanishing anticommutation relations between the modes of 
$\psi^{k}_{\pm}$ are
\begin{align}
\{\tilde{\psi}^{k}_{r},\tilde{\psi}^{l}_{s}\}
=\{\psi^{k}_{r},\psi^{l}_{s}\}=\delta^{kl}\delta_{r-s}.
\end{align}
In this paper $\delta_{m+n}$ and $\delta_{r\pm s}$ denote the Kronecker delta,
namely $\delta_{m+n,0}$ and $\delta_{r\pm s,0}$.
All the other (anti)commutators between the modes vanish. 
We confirm in the next section that these (anti)commutation relations of the modes
are consistent with the canonical anticommutation relations of bosonic fields 
and fermionic fields. 

Here we mention important notice. 
They are free mode representations but they are not 
what we call free field representations in the conformal field theory. 
Because $x^{-}$ which is the zero mode of $X^{-}_{0}$ does not commute 
the twist factor $\hat{\mu}$ which has the momentum $p^{+}$, 
the commutation relations between 
the fields $X^{-}_{0}$ and $f$, $g$, $\lambda_{\pm}$ 
and their Hermitian conjugate fields do not vanish. 
We cannot remove these quantum interaction, 
however $x^{-}$ is absent in the fields $\partial_{\pm}X^{-}_{0}$,
so that $\partial_{\pm}X^{-}_{0}$ are just the same as free fields.
Therefore we can calculate the super-Virasoro algebra without problem.

Finally we write up the definition of the normal orderings about the modes of 
$f,\ g, \lambda_{\pm}$.

$\bullet\,\,$ The normal orderings of $A_{n}^{\dagger}A_{n}$ 
and $B_{n}^{\dagger}B_{n}$ are 
\begin{align}
:A^{\dagger}_{n}A_{n}:&=\left\{
\begin{array}{ll}
A^{\dagger}_{n}A_{n} & \; (n>\hat{\mu})\,,\\
A_{n}A^{\dagger}_{n} & \; (n<\hat{\mu})\,,
\end{array}
\right.  
\label{Normal-order-A}\\
:B^{\dagger}_{n}B_{n}:&=\left\{
\begin{array}{ll}
B^{\dagger}_{n}B_{n} & \; (n>-\hat{\mu})\,, \\
B_{n}B^{\dagger}_{n} & \; (n<-\hat{\mu})\,,
\end{array}
\right. 
\label{Normal-order-B}
\end{align}

$\bullet\,\,$ The normal orderings of 
$\tilde{\lambda}_{r}^{\dagger}\tilde{\lambda}_{r}$
 and $\lambda^{\dagger}_{r}\lambda_{r}$ are 
\begin{align}
:\tilde{\lambda}^{\dagger}_{r}\tilde{\lambda}_{r}:&=\left\{
\begin{array}{ll}
\ \ \tilde{\lambda}^{\dagger}_{r}\tilde{\lambda}_{r} & \; (r>\hat{\mu})\,,\\
-\tilde{\lambda}_{r}\tilde{\lambda}^{\dagger}_{r} & \; (r<\hat{\mu})\,,
\end{array}
\right.  
\label{Normal-order-tildelambda}\\
:\lambda^{\dagger}_{r}\lambda_{r}:&=\left\{
\begin{array}{ll}
\ \ \lambda^{\dagger}_{r}\lambda_{r} & \; (r>-\hat{\mu})\,,\\
-\lambda_{r}\lambda^{\dagger}_{r} & \; (r<-\hat{\mu})\,,
\end{array}
\right. 
\label{Normal-order-lambda}
\end{align}
Here, we assume that $\hat{\mu}$ is a real number excluding all integers 
and all half integers.
Of course, the normal orderings of the modes
$\tilde{\alpha}^{\pm}_{n},\,\alpha^{\pm}_{n},\,
\tilde{\alpha}^{k}_{n}$ and $\alpha^{k}_{n}$
are exactly the same as that in the usual case of free fields.
The normal ordering plays an important role in the calculation of
the anomaly of the super-Virasoro algebra given in \S 7.


\section{Proof of the free-mode representation}

In this section we prove that the general solutions appearing in \S 3 and
the free-mode representations appearing
in \S 5 satisfy the canonical (anti)commutation relations 
for all the covariant string coordinates 
and all the covariant fermionic fields.
Because we need the canonical momentum to quantize the string coordinates,
we obtain the canonical momentum from the action (\ref{action-X})
using $P_{\mu}=\frac{\partial S_{\rm M}}{\partial(\partial_{\tau}X^{\mu})}$,
$P_{Z}=\frac{\partial S_{\rm M}}{\partial(\partial_{\tau}Z)}$
and $P_{Z^{*}}=\frac{\partial S_{\rm M}}{\partial(\partial_{\tau}Z^{*})}$:
\begin{align}
P_{+}&=-\frac{1}{2\pi\alpha'}
\Bigl[
\partial_{\tau}X^{-}+\frac{i\mu}{2}
(Z\partial_{\sigma}Z^{*}
-Z^{*}\partial_{\sigma} Z)+\mu^{2}\partial_{\tau}X^{+}Z^{*}Z\nonumber\\
&\ \ \ \ \ \ \ \ \ \ \ \ \ \ \ 
-\frac{\mu}{2}(\psi^{Z*}_{+}\psi^{Z}_{+}-\psi^{Z*}_{-}\psi^{Z}_{-})
+\frac{i\mu^{2}}{2}\psi^{+}_{+}(Z\psi^{Z*}_{+}+Z^{*}\psi^{Z}_{+})
+\frac{i\mu^{2}}{2}\psi^{+}_{-}(Z\psi^{Z*}_{-}+Z^{*}\psi^{Z}_{-})
\Bigr],\\
P_{-}&=-\frac{1}{2\pi\alpha'}\partial_{\tau}X^{+}\,,\\
P_{Z}&=\frac{1}{4\pi\alpha'}
\bigl[
\partial_{\tau}Z^{*}-i\mu\partial_{\sigma}X^{+}Z^{*}
+\mu(\psi^{+}_{+}\psi^{Z*}_{+}-\psi^{+}_{-}\psi^{Z*}_{-})
\bigr], \\
P_{Z^{*}}&=\frac{1}{4\pi\alpha'}
\bigl[
\partial_{\tau}Z+i\mu\partial_{\sigma}X^{+}Z
-\mu(\psi^{+}_{+}\psi^{Z}_{+}-\psi^{+}_{-}\psi^{Z}_{-})
\bigr]\,,\\
P_{k}&=\frac{1}{2\pi\alpha'}\partial_{\tau}X^{k}\,.
\end{align}
Here we note that it is not necessary to consider the momenta of 
the fermionic fields $\pi_{\mu A}$ by the reason from \S 4.
Although the momentum appears complicated, it can be put into a simpler form
by using the fields 
$X^{+}, \, \tilde{X}^{+}, \,X^{-}_{0}, \,X^{k}, \,f, \ g, \,\psi^{+}_{\pm}$ 
and $\lambda_{\pm}$ which appear in the free-mode representation 
discussed in \S 5.
The field $P_{+}$ becomes the most simplified:
\begin{align}
P_{+}&=-\frac{1}{2\pi\alpha'}\partial_{\tau}X^{-}_{0}, 
\label{P+0}\\
P_{-}&=-\frac{1}{2\pi\alpha'}\partial_{\tau}X^{+}\,,\\
\qquad \qquad \qquad \:\:\:\:\:\:\:
P_{Z}&=\frac{1}{4\pi\alpha'}e^{i\mu\tilde{X}^{+}}
\bigl[
\partial_{+}f^{*}+\partial_{-}g^{*}
+\mu(\psi^{+}_{+}\lambda^{*}_{+}-\psi^{+}_{-}\lambda^{*}_{-})
\bigr]\,,\\
P_{Z^{*}}&=\frac{1}{4\pi\alpha'}e^{-i\mu\tilde{X}^{+}}
\bigl[
\partial_{+}f+\partial_{-}g
-\mu(\psi^{+}_{+}\lambda_{+}-\psi^{+}_{-}\lambda_{-})
\bigr]\,,\\
P_{k}&=\frac{1}{2\pi\alpha'}\partial_{\tau}X^{k}\,.
\end{align} 
Let us remember the canonical (anti)commutation relations 
(\ref{canonical-com-rel:XP})-(\ref{canonical-com-rel:PZ*psi-}) 
in \S 4 for quantization.

First, let us explain almost self-evident parts of the proof.
\vskip 3mm
\begin{enumerate}
\item The proof of the canonical commutation relations between bosonic fields 
without $X^{-}$, $P_{Z}$ and $P_{Z^{*}}$ is the same as the proof 
in our previous paper\cite{Chi-Ya}. 
The reason why we remove $X^{-}$, $P_{Z}$ and $P_{Z^{*}}$ is 
that these fields contain fermionic fields.  
$X^{+}$, $X^{k}$, and $P_{k}$ are trivially free fields. 
so that these fields commute with all the other bosonic fields and fermionic fields. 
$P_{-}$ commutes with all the fields without $X^{-}_{0}$ in $X^{-}$.
Therefore $P_{-}$ and $X^{-}$ satisfy the usual canonical commutation relation. 
Moreover $Z$ and $Z^{*}$ commute with all the fermionic fields. 
So they commute with the fermionic part of $X^{-}$, $P_{Z}$ and $P_{Z^{*}}$. 
Thus $Z$, $Z^{*}$, $X^{-}$, $P_{Z}$ and $P_{Z^{*}}$ satisfy 
the usual canonical commutation relations, and moreover,
$P_{+}$ commutes with $X^{-}$, $P_{Z}$, $P_{Z^{*}}$, $\psi^{Z}_{A}$, $\psi^{Z*}_{A}$ 
and $\psi^{-}_{A}$ because $P_{+}$ commutes with $\tilde{X}^{+}$, 
in the light of the previous paper\cite{Chi-Ya}. 

\vskip 2mm
\item The canonical commutation relations about $\pi_{\mu A}$ are the same as 
those of $\psi^{\mu}_{A}$ through the relation 
$\pi_{\mu A}=\frac{i}{4\pi\alpha'}G_{\mu\nu}\psi^{\nu}_{A}$. 
Because $G_{\mu\nu}$ is the function of $X^{\mu}$,
$G_{\mu\nu}$ commute with $X^{\mu}$ trivially. 
So the canonical commutation relations between $\pi_{\mu A}$ and $X^{\mu}$ are 
the same as the canonical commutation relations between $\psi^{\mu}_{A}$ 
and $X^{\mu}$. 
Moreover we can calculate the commutation relations 
between $\pi_{\mu A}$ and $P_{\mu}$ using the Leibniz rule 
and the commutation relations between $P_{\mu}$ and $G_{\mu\nu}$ 
whose commutation relation is calculated by the canonical commutation relations 
between $X^{\mu}$ and $P_{\mu}$. 
Moreover $G_{\mu\nu}$ commute with $\psi^{\mu}_{A}$. 
Thus we can prove the commutation relations about $\pi_{\mu A}$, 
using  the canonical commutation relations about $\psi^{\mu}_{A}$.

\vskip 2mm
\item Clearly, $\psi^{k}_{A}$ are the same as the ordinary free fermionic fields. 
Thus $\psi^{k}_{A}$ satisfy the usual canonical anticommutation relations 
and $\psi^{k}_{A}$ (anti)commute with all the other fields.

\vskip 2mm
\item The commutation relations between $\psi^{+}_{A}$ and $\psi^{-}_{A}$ 
are reduced to the commutation relation between $\psi^{+}_{A}$ and $\psi^{\ -}_{0A}$ 
because $\psi^{+}_{A}$ (anti)commute with all the other fields.

\vskip 2mm
\item We have constructed the free-mode representations from 
the canonical commutation relation between $X^{-}$ and $\psi^{Z}_{A}$ in \S 5,
so that these relations and their Hermitian conjugate relations are satisfied trivially. 
Moreover $X^{-}$ commutes with $\psi^{-}_{A}$ because $X^{-}$ commutes 
with $Z$, $Z^{*}$, $\psi^{Z}_{A}$, $\psi^{Z*}_{A}$ and $\psi^{+}_{+}$.

\end{enumerate}
\vskip 3mm

We present all the important parts of the proof in next subsections.


\subsection{Canonical anticommutation relations between $\psi^{\mu}_{A}$} 
In this subsection we prove the canonical anticommutation relations between 
$\psi^{\mu}_{A}$. 
First we prove the canonical anticommutation relations between 
$\psi^{Z}_{A}$ and $\psi^{Z*}_{A}$. 
Second we prove the canonical anticommutation relations between 
$\psi^{-}_{A}$ and $\psi^{Z}_{A}\ (\psi^{Z*}_{A})$. 
Finally we prove the canonical anticommutation relations between $\psi^{-}_{A}$. 
The other anticommutation relations between $\psi^{\mu}_{A}$ 
have already been proved in the previous subsection.


\subsubsection{Canonical commutation relations between 
$\psi^{Z}_{A}$ and $\psi^{Z*}_{A}$}
In the proof of the canonical anticommutation relations, 
we had better use the Eqs.(\ref{general_sol:psi^Z}) and (\ref{general_sol:psi^Z*}) 
which are represented by using the fields $Z$ and $Z^{*}$. 
Of course, we can also prove them by using the fields $f$ and $g$.
First we can prove that $\lambda_{\pm}(\sigma^{\pm})$ commutes 
with $\psi^{+}_{\pm}$, $\tilde{X}^{+}$, $Z$ and $Z^{*}$ 
from the canonical anticommutation relations,
so that the anticommutators between $\psi^{Z}_{\pm}$ and $\psi^{Z*}_{\pm}$ 
become the product of 
$e^{-i\mu[\tilde{X}^{+}(\sigma)-\tilde{X}^{+}(\sigma')]}$
and the simple anticommutators between $\lambda_{\pm}$ 
and $\lambda^{*}_{\pm}$, namely the anticommutators 
$\{\psi^{Z}_{\pm}(\tau,\sigma),\psi^{Z*}_{\pm}(\tau,\sigma')\}$ are
\begin{align}
\{\psi^{Z}_{\pm}(\tau,\sigma),\psi^{Z*}_{\pm}(\tau,\sigma')\}
=e^{-i\mu[\tilde{X}^{+}(\sigma)-\tilde{X}^{+}(\sigma')]}
\{\lambda_{\pm}(\sigma^{+}),\lambda^{*}_{\pm}({\sigma'}^{+})\}\,.
\end{align}
Let us pay attention to the case of the anticommutators
between $\lambda_{+}$ and $\lambda^{*}_{+}$.
Substituting the mode expansions of $\lambda_{+}$ and $\lambda^{*}_{+}$, 
this anticommutator becomes
\begin{align}
\{\psi^{Z}_{+}(\tau,\sigma),\psi^{Z*}_{+}(\tau,\sigma')\}
&=2\alpha'e^{-i\mu[\tilde{X}^{+}(\sigma)-\tilde{X}^{+}(\sigma')]}
\sum_{r,s\in{\mathbb Z}+\varepsilon}\{\tilde{\lambda}_{r},
\tilde{\lambda}^{\dagger}_{s}\}
e^{-i(r-\hat{\mu})\sigma^{+}}e^{i(s-\hat{\mu}){\sigma'}^{+}}
\nonumber\\
&=4\pi\alpha'e^{-i[\mu\tilde{X}^{+}(\sigma)-\hat{\mu}\sigma^{+}]}
e^{i[\mu\tilde{X}^{+}(\sigma')-\hat{\mu}{\sigma'}^{+}]}
\times\frac{1}{2\pi}\sum_{r\in{\mathbb Z}+\varepsilon}e^{-ir(\sigma-\sigma')}\,,
\end{align}
where $\{\tilde{\lambda}_{r},\tilde{\lambda}^{\dagger}_{s}\}=\delta_{r-s}$,
and the last term is corresponding to the delta-function.
We must note that the anti-periodic delta-function arises 
when $\lambda_{+}$ and $\lambda^{*}_{+}$ are the NS fermions.
Using the delta-functions and the periodic function
$F(\sigma)=-i[\mu\tilde{X}^{+}(\sigma)-\hat{\mu}\sigma^{+}]$,
the anticommutator becomes
\begin{align}
\{\psi^{Z}_{+}(\tau,\sigma),\psi^{Z*}_{+}(\tau,\sigma')\}
=
\begin{cases}
4\pi\alpha'e^{F(\sigma)-F(\sigma')}\delta(\sigma-\sigma')
\qquad : \text{R sector},\\
4\pi\alpha'e^{F(\sigma)-F(\sigma')}\tilde{\delta}(\sigma-\sigma')
\qquad : \text{NS sector}.
\end{cases}
\label{RNS}
\end{align}
When we treat the case of the R sector,
the function $e^{F(\sigma)-F(\sigma')}$ becomes $1$ 
using the property of the periodic delta-function
because this function is a periodic function of $\sigma$.
When we treat the case of the NS sector, 
we use the property of the anti-periodic delta-function as follows:
\begin{align}
[F(\sigma)-F(\sigma')]\tilde{\delta}(\sigma-\sigma')=0\,,
\label{F-Fdelta}
\end{align}
where we can prove this relation (\ref{F-Fdelta}) by using the Fourier expansion. 
Moreover Taylor expansion of the function $e^{F(\sigma)-F(\sigma')}$ is
$1+\sum_{n=1}^{\infty}\frac{1}{n!}[F(\sigma)-F(\sigma')]^{n}$, 
so that this function also becomes 1 in the case of the NS sector of 
Eq.(\ref{RNS}). 
Therefore the canonical anticommutation relation 
(\ref{canonical-anti-com-rel:psiZpsiZ*}) is proved, 
because we can also prove the case of $\psi^{Z}_{-}$ and $\psi^{Z*}_{-}$ 
using the same procedure.


\subsubsection{Canonical anticommutation relations between 
$\psi^{Z}_{\pm}$ $(\psi^{Z*}_{\pm})$ and $\psi^{-}_{\pm}$}

First we prove the canonical anticommutator relation
$\{\psi^{Z}_{\pm}(\tau,\sigma),\psi^{-}_{\pm}(\tau,\sigma')\}=0$. 
Using the general solutions 
(\ref{general_sol:psi^Z}), (\ref{general sol:psi-+}), (\ref{general sol:psi--}) 
and the canonical anticommutation relations (\ref{canonical-anti-com-rel:psi+psi-}), 
(\ref{canonical-anti-com-rel:psiZpsiZ*}), 
the anticommutator are reduced to 
\begin{align}
\{\psi^{Z}_{\pm}(\tau,\sigma),\psi^{-}_{+}(\tau,\sigma')\}
&=\mp i\mu\{\psi^{+}_{\pm}(\sigma^{\pm}),\psi^{-}_{\pm}(\tau,\sigma')\}
Z(\sigma)\mp\frac{i}{2}\mu Z(\sigma')\{\psi^{Z}_{\pm}(\tau,\sigma),
  \psi^{Z*}_{\pm}(\tau,\sigma')\}\nonumber\\
&=
\begin{cases}
\pm 2\pi i\alpha'[Z(\sigma)-Z(\sigma')]\delta(\sigma-\sigma')
\qquad: \text{R sector},\\
\pm 2\pi i\alpha'[Z(\sigma)-Z(\sigma')]\tilde{\delta}(\sigma-\sigma')
\qquad: \text{NS sector}.
\end{cases}
\label{PsiZ-PsiZ}
\end{align}
Here $Z(\sigma)$ is the periodic function of $\sigma$, 
so that we can use the relation (\ref{F-Fdelta}) in the NS sector again. 
The usual property of the periodic delta-function can also be used 
in the R sector.
Therefore the anticommutators become zero. 
We can prove the case of $\psi^{Z*}_{\pm}$ using the same procedure 
or taking the Hermitian conjugate of $\psi^{Z}_{\pm}$. 

In this proof, we can obtain the useful formulae from Eq.(\ref{PsiZ-PsiZ}).
First, from the canonical anticommutation relations between 
$\psi^{-}_{ \pm}(\tau,\sigma)$ and $\psi^{+}_{\pm}({\sigma'}^{\pm})$,
the following relations are given:
\begin{align}
 \{\psi^{\ -}_{0\pm}(\sigma^{\pm}),\psi^{+}_{\pm}({\sigma'}^{\pm})\}=
 \begin{cases}
 -2\pi\alpha'\delta(\sigma-\sigma') 
 \qquad \qquad\ \ \,  \text{: R sector}\,,\\
 -2\pi\alpha'\tilde{\delta}(\sigma-\sigma')
 \qquad \qquad\ \ \, \text{: NS sector},
 \end{cases}\label{ComRel:psi-0psi+}
 \end{align}
because $\psi^{+}_{\pm}$ (anti)commute with all the other fields.
Second, from the canonical anticommutation relations between 
$\psi^{Z}_{\pm}(\tau,\sigma)$, $\psi^{Z*}_{\pm}(\tau,\sigma)$ 
and $\psi^{-}_{\pm}(\tau,\sigma')$, 
the following relations are given:
\begin{align}
\{\psi^{Z}_{\pm}(\tau,\sigma),\psi^{\ -}_{0\pm}({\sigma'}^{\pm})\}
&=
\begin{cases}
\pm i2\pi\alpha'\mu Z\delta(\sigma-\sigma')
\qquad \quad\ \ \text{: R sector}\,,\\
\pm i2\pi\alpha'\mu Z\tilde{\delta}(\sigma-\sigma')
\qquad \quad\ \ \text{: NS sector}\,,
\end{cases}\label{ComRel:psiZpsi-0}\\
\{\psi^{Z*}_{\pm}(\tau,\sigma),\psi^{\ -}_{0\pm}({\sigma'}^{\pm})\}
&=
\begin{cases}
\mp i2\pi\alpha'\mu Z^{*}\delta(\sigma-\sigma')
\qquad \quad \ \text{: R sector}\,,\\
\mp i2\pi\alpha'\mu Z^{*}\tilde{\delta}(\sigma-\sigma')
\qquad \quad \ \text{: NS sector}\,,
\end{cases}\label{ComRel:psiZ*psi-0}
\end{align}
because the fields $\psi^{Z}_{\pm}$ and $\psi^{Z*}_{\pm}$ have $\psi^{+}_{\pm}$ 
which do not anticommute with $\psi^{\ -}_{0\pm}$.
These formulae are used in the next subsection (6.1.3).


\subsubsection{Canonical anticommutation relations between $\psi^{-}_{A}$}
These anticommutation relations, which are shown in 
the Eq.(\ref{canonical-anti-com-rel:psi-psi-}), 
are not zero in the {\it pp}-wave background.
Let us calculate the anticommutator 
$\{\psi^{-}_{\pm}(\tau,\sigma),\psi^{-}_{\pm}(\tau,\sigma')\}$,
where the general operator solutions of $\psi^{-}_{\pm}$ 
are presented in Eqs.(\ref{general sol:psi-+}) and (\ref{general sol:psi--}). 
The free field $\psi^{\ -}_{0\pm}$ (anti)commute with all the fields without 
$\psi^{+}_{\pm}$, $\psi^{Z}_{\pm}$ and $\psi^{Z*}_{\pm}$. 
Thus the following fields survive in the anticommutators
$\{\psi^{-}_{\pm}(\tau,\sigma),\psi^{-}_{\pm}(\tau,\sigma')\}$:
\begin{align}
\{\psi^{-}_{\pm}(\tau,\sigma),\psi^{-}_{\pm}(\tau,\sigma')\}
=&\pm\frac{i}{2}\mu Z(\sigma')\{\psi^{\ -}_{0\pm}(\sigma^{\pm}),
\psi^{Z*}_{\pm}(\sigma')\}\pm\frac{i}{2}\mu Z^{*}(\sigma')\{\psi^{\ -}_{0\pm}(\sigma^{\pm}),\psi^{Z}_{\pm}(\sigma')\}\nonumber\\
&-\mu^{2}\{\psi^{\ -}_{0\pm}(\sigma^{\pm}),\psi^{+}_{\pm}({\sigma'}^{\pm})\}
Z^{*}(\sigma')Z(\sigma')\nonumber\\
&\mp\frac{i}{2}\mu Z(\sigma)\{\psi^{Z*}_{\pm}(\sigma),\psi^{\ -}_{\pm}({\sigma'}^{\pm})\}+\frac{1}{4}\mu^{2}Z(\sigma)Z^{*}(\sigma')
\{\psi^{Z*}_{\pm}(\sigma),
\psi^{Z}_{\pm}(\sigma')\}\nonumber\\
&\pm\frac{i}{2}\mu Z^{*}(\sigma)\{\psi^{Z}_{\pm}(\sigma),\psi^{\ -}_{\pm}({\sigma'}^{\pm})\}+\frac{1}{4}\mu^{2}Z^{*}(\sigma)Z(\sigma')
\{\psi^{Z}_{\pm}(\sigma),
\psi^{Z*}_{\pm}(\sigma')\}\nonumber\\
&-\mu^{2}\{\psi^{+}_{\pm}(\sigma^{\pm}),\psi^{\ -}_{0\pm}({\sigma'}^{\pm})\}
Z^{*}(\sigma)Z(\sigma).
\end{align}
Using the useful formulae (\ref{ComRel:psi-0psi+})-(\ref{ComRel:psiZ*psi-0}), 
we can reproduce the canonical anticommutation relation 
(\ref{canonical-anti-com-rel:psi-psi-}).
 
We have to note that we can also identify the world-sheet coordinates of 
the periodic function $F(\sigma)$ in the case of anti-periodic delta-function 
$\tilde{\delta}(\sigma-\sigma')$ from the relation (\ref{F-Fdelta}). 
Thus 
$F(\sigma')\tilde{\delta}(\sigma-\sigma')=F(\sigma)\tilde{\delta}(\sigma-\sigma')$.
Moreover we have to take notice of $\{\psi^{\ -}_{0\pm}(\sigma^{\pm}),
\psi^{\ -}_{0\pm}({\sigma'}^{\pm})\}=0$, which means that 
$\psi^{\ -}_{0\pm}$ are just free fields. 


\subsection{Canonical commutation relations between $P_{\mu}$ 
and $\psi^{\mu}_{A}$}
In this subsection we prove the canonical commutation relations between 
$P_{\mu}$ and $\psi^{\mu}_{A}$. 
First we prove the canonical commutation relations between 
$P_{Z}$ $(P_{Z^{*}})$ and $\psi^{Z}_{\pm}$ $(\psi^{Z*}_{\pm})$. 
Second we prove the canonical commutation relations between 
$P_{Z}\,, P_{Z^{*}}$ and $\psi^{-}_{\pm}$. 
The other canonical commutation relations between 
$P_{\mu}$ and $\psi^{\mu}_{A}$ 
have already been proved in the previous subsection.


\subsubsection{Canonical commutation relations between 
$P_{Z}$ $(P_{Z^{*}})$ and $\psi^{Z}_{A}$ $(\psi^{Z*}_{A})$}  
We calculate the commutation relations 
$[P_{Z}(\tau,\sigma),\psi^{Z}_{\pm}(\tau,\sigma')]$. 
First we divide $P_{Z}$ ($P_{Z^{*}}$) into the bosonic part 
$P^{\rm B}_{Z}$ ($P^{\rm B}_{Z^{*}}$) and the fermionic part 
$P^{\rm F}_{Z}$ ($P^{\rm F}_{Z^{*}}$) as follows:
\begin{align}
P_{Z}&=P^{\rm B}_{Z}+P^{\rm F}_{Z},\\
P^{\rm B}_{Z}&=\frac{1}{4\pi\alpha'}\left[\partial_{+}Z^{*}+\partial_{-}Z^{*}
-i\mu(\partial_{+}X^{+}-\partial_{-}X^{+})Z^{*}\right],\\
P^{\rm F}_{Z}&=\frac{\mu}{4\pi\alpha'}[\psi^{+}_{+}\psi^{Z*}_{+}
-\psi^{+}_{-}\psi^{Z*}_{-}]\,,
\end{align}
where $P_{Z^{*}}$ is obtained by Hermitian conjugate of $P_{Z}$.
Therefore the commutators between $P_{Z}$ and $\psi^{Z}_{\pm}$ 
become as follows:
\begin{align}
[P_{Z}(\tau,\sigma),\psi^{Z}_{\pm}(\tau,\sigma')]
=[P^{\rm B}_{Z}(\sigma)+P^{\rm F}_{Z}(\sigma),\psi^{Z}_{\pm}(\sigma')]
=[P^{\rm B}_{Z}(\sigma),\psi^{Z}_{\pm}(\sigma')]
+[P^{\rm F}_{Z}(\sigma),\psi^{Z}_{\pm}(\sigma')]\label{ComRel:PZpsiZ}.
\end{align}
We have to note that $\psi^{Z}_{\pm}$ have the field $Z$ 
in Eq. (\ref{general_sol:psi^Z}), 
so that the commutation relations become as follows:
\begin{align}
[P^{\rm B}_{Z}(\tau,\sigma),\psi^{Z}_{\pm}(\tau,\sigma')]
&=\mp i\mu\psi^{+}_{\pm}({\sigma'}^{\pm})[P^{\rm B}_{Z}(\sigma),Z(\sigma')]
\nonumber\\
&=\mp\mu\psi^{+}_{\pm}({\sigma'}^{\pm})\delta(\sigma-\sigma').
\label{ComRel:PBpsiZ}
\end{align}
Moreover the commutation relations between $P^{\rm F}_{Z}$ and $\psi^{Z}_{\pm}$ 
become as follows:
\begin{align}
[P^{\rm F}_{Z}(\sigma),\psi^{Z}_{\pm}(\sigma')]=
\begin{cases}
\pm\mu\psi^{+}_{\pm}(\sigma^{\pm})\delta(\sigma-\sigma')
\qquad \text{: R sector}\,,\\
\pm\mu\psi^{+}_{\pm}(\sigma^{\pm})\tilde{\delta}(\sigma-\sigma')
\qquad \text{: NS sector}\,.\\
\end{cases}
\end{align}
In the case of the R sector, we can identify the world-sheet coordinates, 
so that  the commutator (\ref{ComRel:PZpsiZ}) vanish. 
In the case of the NS sector, using the relation (\ref{Rel-NS-delta}), 
the commutator (\ref{ComRel:PBpsiZ}) become as follows:
\begin{align}
[P^{\rm B}_{Z}(\tau,\sigma),\psi^{Z}_{\pm}(\tau,\sigma')]
=\mp\mu\psi^{+}_{\pm}({\sigma}^{\pm})\tilde{\delta}(\sigma-\sigma').
\end{align}
Therefore the commutators (\ref{ComRel:PZpsiZ}) also vanish in the NS sector, 
so that the canonical commutation relation 
$[P^{\rm B}_{Z}(\tau,\sigma),\psi^{Z}_{\pm}(\tau,\sigma')]=0$ is proved.


\subsubsection{Canonical commutation relations between 
$P_{Z},\ P_{Z^{*}}$ and $\psi^{-}_{A}$}
Let us pay attention to these commutation relations which are not zero
in the pp-wave background. 
We calculate the commutator $[P_{Z}(\tau,\sigma),\psi^{-}_{\pm}(\tau,\sigma')]$.
Here $P^{\rm B}_{Z}$ commute with $Z^{*}$ and $\psi^{Z*}_{\pm}$,
and the commutators between $P^{\rm B}_{Z}$ and $\psi^{Z}_{\pm}$ are written in 
Eq. (\ref{ComRel:PBpsiZ}), 
and we have to note that $P^{\rm F}_{Z}$ dose not commute with $\psi^{-}_{\pm}$, 
so that the commutators are reduced to 
\begin{align}
[P_{Z}(\tau,\sigma),\psi^{-}_{\pm}(\tau,\sigma')]
=\mp\frac{i}{2}\mu[P^{\rm B}_{Z}(\sigma),Z(\sigma')]\psi^{Z*}_{\pm}(\sigma')
\pm\frac{i}{2}\mu Z^{*}(\sigma')[P^{\rm B}_{Z}(\sigma),\psi^{Z}_{\pm}(\sigma')]
\nonumber\\
-\mu^{2}\psi^{+}_{\pm}Z^{*}(\sigma')[P^{\rm B}_{Z}(\sigma),Z(\sigma')]
+[P^{\rm F}_{Z}(\sigma),\psi^{-}_{\pm}(\sigma')].
\end{align}
Here the commutators
$[P^{\rm F}_{Z}(\sigma),\psi^{-}_{\pm}(\sigma')]$ are calculated as 
\begin{align}
[P^{\rm F}_{Z}(\sigma),\psi^{-}_{\pm}(\sigma')]
=\mp\frac{\mu}{4\pi\alpha'}\psi^{Z}_{\pm}(\sigma)
\{\psi^{+}_{\pm}(\sigma),\psi^{-}_{\pm}(\sigma')\}.
\end{align}
Thus the commutators become the following form:
\begin{align}
[P_{Z}(\tau,\sigma),\psi^{-}_{\pm}(\tau,\sigma')]
=\left(\mp\frac{i}{2}\mu\psi^{Z*}_{\pm}(\sigma')-\frac{\mu^{2}}{2}
Z^{*}\psi^{+}_{\pm}(\sigma')\right)[P^{\rm B}_{Z}(\sigma),Z(\sigma')]
\mp\frac{\mu}{4\pi\alpha'}\psi^{Z}_{\pm}(\sigma)\{\psi^{+}_{\pm}(\sigma),
\psi^{-}_{\pm}(\sigma')\}
\end{align}
Here we note that the delta-function from
$[P_{Z}(\sigma),Z(\sigma')]$ is the periodic delta-function, 
and in the case of the R fermions, we obtain the periodic delta-function 
$\{\psi^{+}_{\pm}(\sigma),\psi^{-}_{\pm}(\sigma')\}
=-2\pi\alpha'\delta(\sigma-\sigma')$. 
So we can calculate it easily and we obtain the canonical commutation relation. 
In the case of the NS fermions, we can obtain the anti-periodic delta-function $\{\psi^{+}_{\pm}(\sigma),\psi^{-}_{\pm}(\sigma')\}
=-2\pi\alpha'\tilde{\delta}(\sigma-\sigma')$, 
and we use the identity (\ref{Rel-NS-delta}) for the NS fermions.
Therefore we obtain the same result as the R fermions, so that we can confirm 
the canonical commutation relation (\ref{canonical-com-rel:PZpsi-}).
We can also prove the commutation relation (\ref{canonical-com-rel:PZ*psi-}) 
using the same procedure.


\subsection{Canonical commutation relations between the bosonic fields}
In this subsection we prove the commutation relations between the bosonic fields
and their momenta, in particular $X^{-}$, $P_{Z}$ and $P_{Z^{*}}$ 
which have fermionic fields. 
First we prove the commutation relation between $P_{Z}$ and $P_{Z^{*}}$. 
Second we prove the commutation relations between $X^{-}$ and $X^{-}$. 
Third we prove the commutation relation between $X^{-}$ and $P_{Z}\ (P_{Z^{*}})$. 
The other commutation relations between the bosonic fields and their momenta 
have already been proved in the previous subsection.


\subsubsection{Canonical commutation relation between $P_{Z}$ and $P_{Z^{*}}$}
We prove $[P_{Z}(\tau,\sigma),P_{Z^{*}}(\tau,\sigma')]=0$.
Since $P_{Z}$ and $P_{Z^{*}}$ contain the fermionic fields,
we divide $P_{Z}$ ($P_{Z^{*}}$) into the bosonic part 
$P^{\rm B}_{Z}$ ($P^{\rm B}_{Z^{*}}$) and the fermionic part 
$P^{\rm F}_{Z}$ ($P^{\rm F}_{Z^{*}}$).
Therefore the commutator is
\begin{align}
[P_{Z}(\tau,\sigma),P_{Z}(\tau,\sigma')]
&=[P^{\rm B}_{Z}(\sigma)+P^{\rm F}_{Z}(\sigma),P^{\rm B}_{Z^{*}}
(\sigma')+P^{\rm F}_{Z^{*}}(\sigma')]\nonumber\\
&=[P^{\rm B}_{Z}(\sigma),P^{\rm F}_{Z^{*}}(\sigma')]
+[P^{\rm F}_{Z}(\sigma),P^{\rm B}_{Z^{*}}(\sigma')]
+[P^{\rm F}_{Z}(\sigma),P^{\rm F}_{Z^{*}}(\sigma')]\,.
\end{align}
Here we proved $[P^{\rm B}_{Z}(\sigma),P^{\rm B}_{Z^{*}}(\sigma')]=0$ 
exactly in the previous paper\cite{Chi-Ya}. 
So we have only to prove $[P^{\rm F}_{Z}(\tau,\sigma)
 ,P^{\rm F}_{Z^{*}}(\tau,\sigma')]=0$ 
because the commutator $[P^{\rm B}_{Z}(\sigma),P^{\rm F}_{Z^{*}}
(\sigma')]$ and $[P^{\rm F}_{Z}(\sigma),P^{\rm B}_{Z^{*}}(\sigma')]$ are trivially zero.
The commutation relation
$[P^{\rm F}_{Z}(\tau,\sigma), P^{\rm F}_{Z^{*}}(\tau,\sigma')]$
becomes as follows:
 \begin{align}
 [P^{\rm F}_{Z}(\tau,\sigma)
 ,P^{\rm F}_{Z^{*}}(\tau,\sigma)]
 =-\left(\frac{\mu}{4\pi\alpha'}\right)^{2}\left(
 [\psi^{+}_{+}(\sigma)\psi^{Z*}_{+}(\sigma),\psi^{+}_{+}(\sigma')\psi^{Z}_{+}(\sigma')]
 +[\psi^{+}_{-}(\sigma)\psi^{Z*}_{-}(\sigma),\psi^{+}_{-}(\sigma')\psi^{Z}_{-}(\sigma')]
 \right)\,.
 \end{align}
 $\psi^{+}_{\pm}$ anticommute with 
 all fermionic fields
 other than $\psi^{\ -}_{0\pm}$,
 so that the commutator is reduced to the following:
 \begin{align}
 [P^{\rm F}_{Z}(\tau,\sigma)
 ,P^{\rm F}_{Z^{*}}(\tau,\sigma)]&=\left(\frac{\mu}{4\pi\alpha'}\right)^{2}
 \left(
 \psi^{+}_{+}(\sigma)\psi^{+}_{+}(\sigma')\{\psi^{Z*}_{+}(\sigma),\psi^{Z*}_{+}(\sigma')\}
 +\psi^{+}_{-}(\sigma)\psi^{+}_{-}(\sigma')\{\psi^{Z*}_{-}(\sigma),\psi^{Z*}_{-}(\sigma')\}
 \right)\,,
 \end{align}
 where 
  $\{\psi^{Z*}_{\pm}(\sigma),\psi^{Z*}_{\pm}(\sigma')\}
  =4\pi\alpha'\delta(\sigma-\sigma')$
 in the R sector and
 $\{\psi^{Z*}_{\pm}(\sigma),\psi^{Z*}_{\pm}(\sigma')\}
 =4\pi\alpha'\tilde{\delta}(\sigma-\sigma')$ 
 in the NS sector.
 Using the relation (\ref{Rel-NS-delta}), we can make the periodic delta-function
 even in the NS sector, so that we identify the world-sheet coordinates of
 $\psi^{+}_{\pm}(\sigma)\psi^{+}_{\pm}(\sigma')$.
 From the canonical anticommutation relations
$\{\psi^{+}_{\pm}(\sigma),\psi^{+}_{\pm}(\sigma')\}=0$,
$(\psi^{+}_{\pm})^{2}$ vanishes.
Therefore the commutation relation
$[P_{Z}(\tau,\sigma),P_{Z^{*}}(\tau,\sigma')]=0$ is proved.

 
\subsubsection{Canonical commutation relation between $X^{-}$ and $X^{-}$}
We calculate the commutator $[X^{-}(\tau,\sigma),X^{-}(\tau,\sigma')]$.
Here we define $X^{-}_{\rm B}$, which is the bosonic part of $X^{-}$,
as $X^{-}_{0}+X^{-}_{\rm 1B}+X^{-}_{2}$. 
We proved $[X^{-}_{\rm B}(\tau,\sigma),X^{-}_{\rm B}(\tau,\sigma')]=0$ 
exactly in the previous paper\cite{Chi-Ya} 
and $[X^{-}_{\rm B}(\tau,\sigma),X^{-}_{\rm 1F}(\tau,\sigma')]$ is trivially zero, 
so that we have only to prove 
$[X^{-}_{\rm 1F}(\tau,\sigma),X^{-}_{\rm 1F}(\tau,\sigma')]=0$,
where $X^{-}_{\rm 1F}(\tau,\sigma)$ is defined in (\ref{X-1F}).
We have only to prove the case of $\tau=0$ 
because we can describe the time evolution using the Hamiltonian of the system
in the same method of the previous paper\cite{Chi-Ya}.
The commutator becomes the following:
\begin{align}
[&X^{-}_{\rm 1F}(\sigma),X^{-}_{\rm 1F}(\sigma')]|_{\tau=0}\nonumber\\
&=-\mu^{2}{\alpha'}^{2}\sum_{m,n\neq 0}\sum_{r,s\in Z+\epsilon}\frac{1}{mn}
[\tilde{\lambda}^{\dagger}_{r}\tilde{\lambda}_{m+r}e^{-im\sigma}
-\lambda^{\dagger}_{r}\lambda_{m+r}e^{im\sigma}
,\tilde{\lambda}^{\dagger}_{s}\tilde{\lambda}_{n+s}e^{-in\sigma'}
-\lambda^{\dagger}_{s}\lambda_{n+s}e^{in\sigma'}]\nonumber\\
&=-\mu^{2}{\alpha'}^{2}\sum_{m,n\neq 0}\sum_{r,s\in Z+\epsilon}\frac{1}{mn}
\left([\tilde{\lambda}^{\dagger}_{r}\tilde{\lambda}_{m+r},
\tilde{\lambda}^{\dagger}_{s}\tilde{\lambda}_{n+s}]e^{-im\sigma-in\sigma'}
+[\lambda^{\dagger}_{r}\lambda_{m+r},\lambda^{\dagger}_{s}
\lambda_{n+s}]e^{im\sigma+in\sigma'}\right).
\end{align}
Here we use the formula of commutation relation of Grassmann numbers:
\begin{align}
[\alpha\beta,\gamma\delta]
=\alpha\{\beta,\gamma\}\delta-\{\alpha,\gamma\}\beta\delta
+\gamma\alpha\{\beta,\delta\}-\gamma\{\alpha,\delta\}\beta\,,
\end{align}
where $\alpha,\ \beta,\ \gamma, \delta$ are arbitrary Grassmann numbers. 
Using this formula, the commutator becomes
\begin{align}
\sum_{r,s\in Z+\epsilon}[\lambda^{\dagger}_{r}\lambda_{m+r},
\lambda^{\dagger}_{s}\lambda_{n+s}]
&=\sum_{r,s\in Z+\epsilon}\left(\lambda^{\dagger}_{r}\lambda_{n+s}\delta_{r-s+m}
-\lambda^{\dagger}_{s}\lambda_{m+r}\delta_{s-r+n}\right)\nonumber\\
&=\sum_{r\in Z+\epsilon}\left(\lambda^{\dagger}_{r}\lambda_{n+m+r}
-\lambda^{\dagger}_{r-n}\lambda_{r+m}\right).
\label{lambda-lambda}
\end{align}
After carrying out the summation over $s$, 
we change the suffix as $r\rightarrow r-n$ in the first term in Eq.(\ref{lambda-lambda}), 
so that this commutator vanish. 
We can prove the case of $\tilde{\lambda}_{r}$ in the same method.
Of course we can prove the case of $\tau\neq 0$. 
Therefore $[X^{-}(\tau,\sigma),X^{-}(\tau,\sigma')]=0$ is proved.


\subsubsection{Canonical commutation relation between $X^{-}$ and $P_{Z}\ (P_{Z^{*}})$}
We prove the commutation relation  $[X^{-}(\tau,\sigma),P_{Z}(\tau,\sigma')]=0$.
We can divide $X^{-}$ into $X^{-}_{\rm B}$ and $X^{-}_{\rm 1F}$. 
and moreover, we can also divide $P_{Z}$ into $P^{\rm B}_{Z}$ and $P^{F}_{Z}$. 
We proved the commutation relation between $X^{-}_{\rm B}$ and $P^{\rm B}_{Z}$ 
in the previous paper\cite{Chi-Ya},
and the commutation relation between $X^{-}_{\rm 1F}$ and $P^{\rm B}_{Z}$ 
is trivially zero. So we have only to prove the commutation relation 
between $X^{-}$ and $P^{\rm F}_{Z}$. 
\begin{align}
[X^{-}(\tau,\sigma),P_{Z}(\tau,\sigma')]=[X^{-}(\tau,\sigma),P^{\rm F}_{Z}(\tau,\sigma')]
\end{align}
where $P^{\rm F}_{Z}$ consists of $\psi^{+}_{\pm}$ and $\psi^{Z*}_{\pm}$.
$\psi^{+}_{\pm}$ commute with $X^{-}$ trivially, 
and we have already proved the canonical commutation relation between $X^{-}$ 
and $\psi^{Z*}_{\pm}$ in the previous section. 
Therefore the commutation relation between $X^{-}$ and $P_{Z}$ is proved. 
We can also prove the case of $P_{Z*}$ using the same methods.
\\

We have thus completed the proofs of all the equal-time canonical 
(anti)commutation relations in the free-mode representations.


\section{Super-Virasoro algebra and anomaly}

In this section we define the energy-momentum tensor, the supercurrent 
and the super-Virasoro generators using the normal procedure. 
It is characteristic that they are represented as almost free cases in using
our general operator solutions and our free mode representation.
Therefore, we can exactly calculate the commutators or the anticommutators
between the super-Virasoro generators and obtain the super-Virasoro anomaly.

The energy momentum tensor of the matter $T^{\rm M}_{\alpha \beta}$
is defined as the response to variations of the world-sheet zweibein
in the action, and the supercurrent of the matter $T^{\rm M}_{{\rm F}\alpha A}$
is similarly defined as the response to variations of the world-sheet gravitino
in the action (\ref{Action-M}) :
\begin{align}
&T^{\rm M}_{\alpha \beta}=-\frac{2\pi}{e}\frac{\delta S_{\rm M}}{\delta e^{\alpha}_{\,\,\,a}}
{e}_{\beta a}\,\,\,,
\,\,\,\,\,\,\,\,\,\,\,\,\,\,\,\,\,\,\,\,\,\,\,\,\,\,\,\,\,\,\,\,
T^{\rm M}_{{\rm F}\alpha A}=-\frac{\sqrt{2}\pi i}{e}
\frac{\delta S_{\rm M}}{\delta \bar{\chi}^{\alpha A}}\,\,\,,
\end{align}
where $A$ is the spinor component, and  the variation of gravitino is
performed from the right-hand side.
Due to the tracelessness of the energy momentum tensor and 
the supercurrent, the only non-vanishing components are
$T^{\rm M}_{\pm\pm}$ and $T^{\rm M}_{{\rm F}\pm\pm}$.
Here, the energy momentum tensor $T^{\rm M}_{\pm\pm}$ are written in 
the world-sheet light-cone coordinates ($\sigma^{\pm}$)  system 
according to the rules of tensor analysis, 
and ${}_{++}\,\,({}_{--})$ in the supercurrent $T^{\rm M}_{{\rm F}++}\,\,
(T^{\rm M}_{{\rm F}--})$
denotes the $\sigma^{+}\,\,(\sigma^{-})$ vector component
and the upper (lower) spinor component.
We fix the covariant gauge as 
$e_{\alpha}^{\,\,\,a}=\delta_{\alpha}^{\,\,\,a}$ and $\chi_{\alpha}=0$
after the variation of the action.
Therefore the energy-momentum tensor and the supercurrent of the matter
in the {\it pp}-wave background with the NS-NS flux are
\begin{align}
T^{\rm M}_{++}&=\frac{1}{\alpha'} 
\Big[-2\partial_{+}X^{+}\partial_{+}X^{-}
-\mu^{2}Z^{*}Z\partial_{+}X^{+}\partial_{+}X^{+}
+\partial_{+}Z^{*}\partial_{+}Z
+\partial_{+}X^{k}\partial_{+}X^{k}
\nonumber\\
&\qquad\quad+\frac{i}{2}
\Big\{-\psi^{+}_{+}\partial_{+}\psi^{-}_{+}-\psi^{-}_{+}\partial_{+}\psi^{+}_{+}
-\mu^{2}Z^{*}Z\psi^{+}_{+}\partial_{+}\psi^{+}_{+}
+\frac{1}{2}\psi^{Z}_{+}\partial_{+}\psi^{Z*}_{+}
+\frac{1}{2}\psi^{Z*}_{+}\partial_{+}\psi^{Z}_{+}
\nonumber\\
&\qquad\quad\qquad\,\,\,\,
-\mu^2\partial_{+}X^{+}Z^{*}\psi^{+}_{+}\psi^{Z}_{+}
-\mu^2\partial_{+}X^{+}Z\psi^{+}_{+}\psi^{Z*}_{+}
-i\mu\partial_{+}Z\psi^{+}_{+}\psi^{Z*}_{+}
+i\mu\partial_{+}Z^{*}\psi^{+}_{+}\psi^{Z}_{+}
\nonumber\\
&\qquad\quad\qquad\,\,\,\,
+\frac{i}{2}\mu\partial_{+}X^{+}\psi^{Z}_{+}\psi^{Z*}_{+}
-\frac{i}{2}\mu\partial_{+}X^{+}\psi^{Z*}_{+}\psi^{Z}_{+}
+\psi^{k}_{+}\partial_{+}\psi^{k}_{+}
\Big\}
\Big]\,\,\,,\\
T^{\rm M}_{--}&=\frac{1}{\alpha'} 
\Big[-2\partial_{-}X^{+}\partial_{-}X^{-}
-\mu^{2}Z^{*}Z\partial_{-}X^{+}\partial_{-}X^{+}
+\partial_{-}Z^{*}\partial_{-}Z
+\partial_{-}X^{k}\partial_{-}X^{k}
\nonumber\\
&\qquad\quad+\frac{i}{2}
\Big\{-\psi^{+}_{-}\partial_{-}\psi^{-}_{-}-\psi^{-}_{-}\partial_{-}\psi^{+}_{-}
-\mu^{2}Z^{*}Z\psi^{+}_{-}\partial_{-}\psi^{+}_{-}
+\frac{1}{2}\psi^{Z}_{-}\partial_{-}\psi^{Z*}_{-}
+\frac{1}{2}\psi^{Z*}_{-}\partial_{-}\psi^{Z}_{-}
\nonumber\\
&\qquad\quad\qquad\,\,\,\,
-\mu^2\partial_{-}X^{+}Z^{*}\psi^{+}_{-}\psi^{Z}_{-}
-\mu^2\partial_{-}X^{+}Z\psi^{+}_{-}\psi^{Z*}_{-}
+i\mu\partial_{-}Z\psi^{+}_{-}\psi^{Z*}_{-}
-i\mu\partial_{-}Z^{*}\psi^{+}_{-}\psi^{Z}_{-}
\nonumber\\
&\qquad\quad\qquad\,\,\,\,
-\frac{i}{2}\mu\partial_{+}X^{+}\psi^{Z}_{-}\psi^{Z*}_{-}
+\frac{i}{2}\mu\partial_{+}X^{+}\psi^{Z*}_{-}\psi^{Z}_{-}
+\psi^{k}_{-}\partial_{-}\psi^{k}_{-}
\Big\}
\Big]\,\,\,,\\
T^{\rm M}_{{\rm F}++}&=\frac{\sqrt{2}}{\alpha'}
\Big[-\psi^{+}_{+}\partial_{+}X^{-}
-\psi^{-}_{+}\partial_{+}X^{+}-\mu^2 Z^{*}Z\psi^{+}_{+}\partial_{+}X^{+}
\nonumber\\
&\qquad\quad\,\,+\frac{1}{2}\psi^{Z}_{+}\partial_{+}Z^{*}
+\frac{1}{2}\psi^{Z*}_{+}\partial_{+}Z
+\psi^{k}_{+}\partial_{+}X^{k}
-\frac{1}{2}\mu\psi^{+}_{+}\psi^{Z}_{+}\psi^{Z*}_{+}
\Big]\,\,\,,\\
T^{\rm M}_{{\rm F}--}&=\frac{\sqrt{2}}{\alpha'}
\Big[-\psi^{+}_{-}\partial_{-}X^{-}
-\psi^{-}_{-}\partial_{-}X^{+}-\mu^2 Z^{*}Z\psi^{+}_{-}\partial_{-}X^{+}
\nonumber\\
&\qquad\quad\,\,+\frac{1}{2}\psi^{Z}_{-}\partial_{-}Z^{*}
+\frac{1}{2}\psi^{Z*}_{-}\partial_{-}Z
+\psi^{k}_{-}\partial_{-}X^{k}
+\frac{1}{2}\mu\psi^{+}_{-}\psi^{Z}_{-}\psi^{Z*}_{-}
\Big]\,\,\,.
\end{align}
Moreover, substituting the general solutions of the matter, 
(\ref{1})-(\ref{X}), 
(\ref{2}),
(\ref{Psi+k}),
(\ref{general_sol:psi^Z}),
(\ref{general_sol:psi^Z*}),  
(\ref{4}),
(\ref{Psi-}) and (\ref{X-}),
into the energy-momentum tensor and the supercurrent, they become
\begin{align}
T_{++}^{\rm M}&=\frac{1}{\alpha'}\Big[-2:\partial_{+}X^{+}\partial_{+}X^{-}_{0}:
+\mu\partial_{+}X^{+}(J_{\rm B}+J_{\rm F})+:\partial_{+}f^{*}\partial_{+}f:
+:\partial_{+}X^{k}\partial_{+}X^{k}: \nonumber\\
&\qquad\quad
+\frac{i}{2}
\Big\{-:\psi^{+}_{+}\partial_{+}\psi^{\ -}_{0+}:
-:\psi^{\ -}_{0+}\partial_{+}\psi^{+}_{+}:
+\frac{1}{2}:\lambda_{+}\partial_{+}\lambda^{*}_{+}:
+\frac{1}{2}:\lambda^{*}_{+}\partial_{+}\lambda_{+}:
+:\psi^{k}_{+}\partial_{+}\psi^{k}_{+}:
\Big\}
\Big]\,\,\,,\\
T_{--}^{\rm M}&=\frac{1}{\alpha'}\Big[-2:\partial_{-}X^{+}\partial_{-}X^{-}_{0}:
-\mu\partial_{-}X^{+}(J_{\rm B}+J_{\rm F})+:\partial_{-}g^{*}\partial_{-}g:
+:\partial_{-}X^{k}\partial_{-}X^{k}: \nonumber\\
&\qquad\quad
+\frac{i}{2}
\Big\{-:\psi^{+}_{-}\partial_{-}\psi^{\ -}_{0-}:
-:\psi^{\ -}_{0-}\partial_{-}\psi^{+}_{-}:
+\frac{1}{2}:\lambda_{-}\partial_{-}\lambda^{*}_{-}:
+\frac{1}{2}:\lambda^{*}_{-}\partial_{-}\lambda_{-}:
+:\psi^{k}_{-}\partial_{-}\psi^{k}_{-}:
\Big\}
\Big]\,\,,\\
T^{\rm M}_{{\rm F}++}&=\frac{\sqrt{2}}{\alpha'}
\Big[-\psi^{+}_{+}\partial_{+}X^{-}_{+}-\psi^{\ -}_{0+}\partial_{+}X^{+}
+\frac{1}{2}\mu\psi^{+}_{+}(J_{\rm B}+J_{\rm F})
+\frac{1}{2}\lambda_{+}\partial_{+}f^{*}
+\frac{1}{2}\lambda^{*}_{+}\partial_{+}f
+\psi^{k}_{+}\partial_{+}X^{k}_{+}
\Big]\,\,,\\
T^{\rm M}_{{\rm F}--}&=\frac{\sqrt{2}}{\alpha'}
\Big[-\psi^{+}_{-}\partial_{-}X^{-}_{-}-\psi^{\ -}_{0-}\partial_{-}X^{+}
-\frac{1}{2}\mu\psi^{+}_{-}(J_{\rm B}+J_{\rm F})
+\frac{1}{2}\lambda_{-}\partial_{-}g^{*}
+\frac{1}{2}\lambda^{*}_{-}\partial_{-}g
+\psi^{k}\partial_{-}X^{k}
\Big]\,\,.
\end{align}
Although the interaction remains in the terms
$\partial_{\pm}X^{+}(J_{\rm B}+J_{\rm F})$ and $\psi^{+}_{\pm}(J_{\rm B}+J_{\rm F})$,
we can completely calculate the anomalies of the super-Virasoro algebra
in the same manner as the free fields.

Next, we define the super-Virasoro generators, which are 
the Fourier coefficients of the energy-momentum tensor
and the supercurrent :
\begin{align}
\tilde{L}^{\rm M}_{n}&=\int_{0}^{2\pi}\frac{d\sigma}{2\pi}
e^{in\sigma}T^{\rm M}_{++},\qquad\quad
L^{\rm M}_{n}=\int_{0}^{2\pi}\frac{d\sigma}{2\pi}e^{-in\sigma}T^{\rm M}_{--}\,, \\
\tilde{G}^{\rm M}_{r}&=\int_{0}^{2\pi}\frac{d\sigma}{2\pi}
e^{ir\sigma}T^{\rm M}_{{\rm F}++},\qquad\quad
G^{\rm M}_{r}=\int_{0}^{2\pi}\frac{d\sigma}{2\pi}
e^{-ir\sigma}T^{\rm M}_{{\rm F}--}\,.
\end{align}
The generator $\tilde{L}^{\rm M}_{n}$  ($L^{\rm M}_{n}$) can be divided 
into $\tilde{L}_{n}^{(+-k)}$ ($L_{n}^{(+-k)}$), 
which is a part of $X^{+}$, $X^{-}_{0}$, $X^{k}$, $\psi^{+}$, $\psi^{-}_{0}$,
$\psi^{k}$, $J_{\rm B}$ and $J_{\rm F}$,
$\tilde{L}^{f}_{n}$ ($L^{g}_{n}$), which is a part of $f$ ($g$), 
and $\tilde{L}^{\lambda}_{n}$ ($L^{\lambda}_{n}$), 
which is a part of $\lambda_{+}$ ($\lambda_{-}$), 
as
$\tilde{L}_{n}^{\rm M}=\tilde{L}_{n}^{(+-k)}+\tilde{L}_{n}^{f}+\tilde{L}_{n}^{\lambda}$
and $L_{n}^{\rm M}=L_{n}^{(+-k)}+L_{n}^{g}+L_{n}^{\lambda}$.
We can analogously divide the generator $\tilde{G}^{\rm M}_{r}$ ($G^{\rm M}_{r}$)
into $\tilde{G}_{r}^{(+-k)}$ ($G_{r}^{(+-k)}$), 
which is a part of $X^{+}$, $X^{-}_{0}$, $X^{k}$, $\psi^{+}$, $\psi^{-}_{0}$,
$\psi^{k}$, $J_{\rm B}$ and $J_{\rm F}$,
$\tilde{G}^{\lambda f}_{r}$ ($G^{\lambda g}_{r}$), which is a part of 
$f$ and $\lambda_{+}$ ($g$ and $\lambda_{-}$),
as 
$\tilde{G}_{r}^{\rm M}=\tilde{G}_{r}^{(+-k)}+\tilde{G}_{r}^{\lambda f}$
and $G_{r}^{\rm M}=G_{r}^{(+-k)}+G_{r}^{\lambda g}$.
Moreover in our free-mode representation, the super-Virasoro generators 
are as follows.\\

$\bullet\,\,$ $\tilde{L}_{n}^{(+-k)}$ and $L_{n}^{(+-k)}$ in
the super-Virasoro generators are
\begin{align}
\tilde{L}^{(+-k)}_{n}&=
\frac{1}{2}\sum_{m\in \mathbb{Z}}
\big[-2:\tilde{\alpha}^{+}_{n-m}\tilde{\alpha}^{-}_{m}:
+:\tilde{\alpha}^{k}_{n-m}\tilde{\alpha}^{k}_{m}:\big]
\nonumber\\
&+\frac{1}{2}
\sum_{r\in \mathbb{Z}+\varepsilon}\big[-2\big( r-\frac{n}{2} \big)
:\tilde{\psi}^{+}_{n-r}\tilde{\psi}^{-}_{r}:
+\big(r-\frac{n}{2}\big):\tilde{\psi}^{k}_{n-r}\tilde{\psi}^{k}_{r}:
\big]
\nonumber\\
&+\frac{\mu}{\sqrt{2\alpha'}}\tilde{\alpha}^{+}_{n}(J_{\rm B}+J_{\rm F})\,, \\
L^{(+-k)}_{n}&=
\frac{1}{2}\sum_{m\in \mathbb{Z}}\big[-2:\alpha^{+}_{n-m}\alpha^{-}_{m}:
+:\alpha^{k}_{n-m}\alpha^{k}_{m}:\big]
\nonumber\\
&+\frac{1}{2}
\sum_{r\in \mathbb{Z}+\varepsilon}\big[-2\big( r-\frac{n}{2} \big)
:\psi^{+}_{n-r}\psi^{-}_{r}:
+\big(r-\frac{n}{2}\big):\psi^{k}_{n-r}\psi^{k}_{r}:
\big] 
\nonumber\\
&-\frac{\mu}{\sqrt{2\alpha'}}\alpha^{+}_{n}(J_{\rm B}+J_{\rm F})\,.
\end{align}  
   
$\bullet\,\,$ $\tilde{L}_{n}^{f}$, $L_{n}^{g}$, $\tilde{L}^{\lambda}_{n}$,
and $L^{\lambda}_{n}$ in the super-Virasoro generators are
\begin{align}
\tilde{L}^{f}_{n}&=
\sum_{m\in \mathbb{Z}}(m-n-\hat{\mu})(m-\hat{\mu}):
\hat{A}^{\dagger}_{m-n}\hat{A}_{m}:\,,\\
L^{g}_{n}&=
\sum_{m\in \mathbb{Z}}(m-n+\hat{\mu})(m+\hat{\mu}):
\hat{B}^{\dagger}_{m-n}\hat{B}_{m}:\,,\\
\tilde{L}^{\lambda}_{n}&=
\sum_{r\in \mathbb{Z}+\varepsilon}\big(r-\frac{n}{2}-\hat\mu\big)
:\tilde{\lambda}^{\dag}_{r-n}\tilde{\lambda}_{r}:\,,\\
L^{\lambda}_{n}&=
\sum_{r\in \mathbb{Z}+\varepsilon}\big(r-\frac{n}{2}+\hat\mu\big)
:\lambda^{\dag}_{r-n}\lambda_{r}:\,.
\end{align}

$\bullet\,\,$ $\tilde{G}_{r}^{(+-k)}$, $G_{r}^{(+-k)}$, 
$\tilde{G}^{\lambda f}_{r}$ and $G^{\lambda g}_{r}$
in the super-Virasoro generators are
\begin{align}
\tilde{G}_{r}^{(+-k)}&=
\sum_{n\in \mathbb{Z}}\big[-\tilde{\psi}^{+}_{r-n}\tilde{\alpha}^{-}_{n}
-\tilde{\psi}^{-}_{r-n}\tilde{\alpha}^{+}_{n}
+\tilde{\psi}^{k}_{r-n}\tilde{\alpha}^{k}_{n}
\big]+\frac{\mu}{\sqrt{2\alpha'}}\tilde{\psi}^{+}_{r}(J_{\rm B}+J_{\rm F})\,,\\
G_{r}^{(+-k)}&=
\sum_{n\in \mathbb{Z}}\big[-\psi^{+}_{r-n}\alpha^{-}_{n}
-\psi^{-}_{r-n}\alpha^{+}_{n}
+\psi^{k}_{r-n}\alpha^{k}_{n}
\big]-\frac{\mu}{\sqrt{2\alpha'}}\psi^{+}_{r}(J_{\rm B}+J_{\rm F})\,,\\
\tilde{G}^{\lambda f}_{r}&=
i\sum_{n\in \mathbb{Z}}(n-\hat\mu)(\tilde{\lambda}_{n+r}\hat{A}^{\dag}_{n}
-\tilde{\lambda}^{\dag}_{n-r}\hat{A}_{n})
\,,\\
G^{\lambda g}_{r}&=
i\sum_{n\in \mathbb{Z}}(n+\hat\mu)(\lambda_{n+r}\hat{B}^{\dag}_{n}
-\lambda^{\dag}_{n-r}\hat{B}_{n})\,.
\end{align}
Here we define $\tilde{\alpha}^{\pm}_{0}=\alpha^{\pm}_{0}
=\sqrt{\frac{\alpha'}{2}}p^{\pm},\,$ 
$\tilde{\alpha}^{k}_{0}=\alpha^{k}_{0}=\sqrt{\frac{\alpha'}{2}}p^{k},\,$
$\hat{A}_{n}=\frac{1}{\sqrt{|n-\hat{\mu}|}}A_{n}$,
$\hat{B}_{n}=\frac{1}{\sqrt{|n+\hat{\mu}|}}B_{n}$,
 and moreover $\hat\mu=\mu\alpha' p^{+}$,
$\varepsilon=0$ in the R sector
and $\varepsilon=\frac{1}{2}$ in the NS sector, as we define in previous section. 

Calculating the commutators between the generators 
$\tilde{L}_{n}^{f}$, $L_{n}^{g}$, $\tilde{L}_{n}^{\lambda}$
and $L_{n}^{\lambda}$, we obtain
\begin{align}
[\tilde{L}^{f}_{m},\tilde{L}^{f}_{n}]&=
(m-n)\tilde{L}^{f}_{m+n}+\tilde{A}^{f}(m)\delta_{m+n}\,,\\
[L^{g}_{m},L^{g}_{n}]&=
(m-n)L^{g}_{m+n}+A^{g}(m)\delta_{m+n}\,,\\
[\tilde{L}^{\lambda}_{m},\tilde{L}^{\lambda}_{n}]&=
(m-n)\tilde{L}^{\lambda}_{m+n}+\tilde{A}^{\lambda}(m)\delta_{m+n}\,,\\
[L^{\lambda}_{m},L^{\lambda}_{n}]&=
(m-n)L^{\lambda}_{m+n}+A^{\lambda}(m)\delta_{m+n}\,,
\end{align}
where $\tilde{A}^{f}(m)$, $A^{g}(m)$, $\tilde{A}^{\lambda}(m)$ 
and $A^{\lambda}(m)$ represent the anomalies of the algebra for
$\tilde{L}_{n}^{f}$, $L_{n}^{g}$, $\tilde{L}^{\lambda}_{m}$ 
and $L^{\lambda}_{m}$ ;
the anomalies are
\begin{align}
\tilde{A}^{f}(m)&=A^{g}(m)=\frac{1}{6}(m^{3}-m)-(\hat{\mu}-[\hat{\mu}])
(\hat{\mu}-[\hat{\mu}]-1)m\,,
\label{anomaly:f-g}\\
\tilde{A}^{\lambda}(m)&=A^{\lambda}(m)=
\begin{cases}
\frac{1}{12}(m^{3}-m)+\frac{1}{4}m
+(\hat{\mu}-[\hat{\mu}])(\hat{\mu}-[\hat{\mu}]-1)m \,\,\,\,\,\,\,\,\,\,
& \text{: \mbox{R sector}}\,,\\
\frac{1}{12}(m^{3}-m)+\big(\hat{\mu}-\big[\hat{\mu}
+\frac{1}{2}\big]\big)^{2}m
\,\,\,\,\,\,\,\,\,\,
\qquad\qquad\quad\,\,\,\,
& \text{: \mbox{NS sector}}\,,
\end {cases}
\end{align}
where $[\hat{\mu}]$ is the greatest integer that is not beyond $\hat{\mu}$,
namely, the Gauss' symbol.
Because the twisted fields $f$ and $g$ are complex fields, and each of them has 
two degrees of freedom, the coefficient $\frac{1}{6}$ appears 
in Eqs.(\ref{anomaly:f-g}).
As a known case, the anomalies of the algebra for $\tilde{L}_{n}^{(+-k)}$ 
and $L_{n}^{(+-k)}$
are
$\tilde{A}^{(+-k)}(m)=A^{(+-k)}(m)=\frac{D-2}{8}(m^{3}-2\varepsilon m)$.
Next, the anticommutators between the generators 
$\tilde{G}^{\lambda f}_{r}$ and $G^{\lambda g}_{r}$ are
\begin{align}
\big\{\tilde{G}^{\lambda f}_{r}, \tilde{G}^{\lambda f}_{s}\big\}&=
2\big(\tilde{L}^{\lambda}_{r+s}+\tilde{L}^{f}_{r+s}\big)+
\tilde{B}^{\lambda f}(r)\delta_{r+s}\,,\\
\big\{G^{\lambda g}_{r}, G^{\lambda g}_{s}\big\}&=
2\big(L^{\lambda}_{r+s}+L^{g}_{r+s}\big)+B^{\lambda g}(r)\delta_{r+s}\,,
\end{align}
where $\tilde{B}^{\lambda f}(r)$ and $B^{\lambda g}(r)$ represent
the anomalies of the algebra for $\tilde{G}^{\lambda f}_{r}$
and $G^{\lambda g}_{r}$ ; the anomalies are
\begin{align}
\tilde{B}^{\lambda f}(r)=B^{\lambda g}(r)=
\begin{cases}
r^{2}
& \text{: \mbox{R sector}}\,,
\\
r^{2}-\frac{1}{4}+\big[\hat{\mu}+\frac{1}{2}\big]^2
-[\hat\mu]^2-[\hat{\mu}]-2\big(\big[\hat{\mu}+\frac{1}{2}\big]
-[\hat{\mu}]-\frac{1}{2}\big)\hat\mu 
& \text{: \mbox{NS sector}}\,.
\end{cases}
\end{align}

When we gather the super-Virasoro generators for the matter,
then the super-Virasoro algebra becomes
\begin{align}
\left[\tilde{L}_{m}^{\rm M},\tilde{L}_{n}^{\rm M}\right]
=\,(m-n)\tilde{L}_{m+n}^{\rm M}+\tilde{A}^{\rm M}(m)\delta_{m+n}\,,
\,\,\,\,
\left[L_{m}^{\rm M},L_{n}^{\rm M}\right]
=(m-n)L_{m+n}^{\rm M}+A^{\rm M}(m)\delta_{m+n}\,,&
\\
\left[\tilde{L}^{\rm M}_{m}, \tilde{G}^{\rm M}_{r}\right]
=\left(\frac{m}{2}-r\right) \tilde{G}^{\rm M}_{m+r}\,,
\qquad\qquad\qquad\,\,\,\,\,
\left[L^{\rm M}_{m}, G^{\rm M}_{r}\right]
=\left(\frac{m}{2}-r\right) G^{\rm M}_{m+r}\,,
\quad\qquad\qquad\quad&
\\
\left\{\tilde{G}^{\rm M}_{r}, \tilde{G}^{\rm M}_{s}\right\}
=2\tilde{L}^{\rm M}_{r+s}+\tilde{B}^{\rm M}(r)\delta_{r+s}\,,
\qquad\qquad\,\,\,
\left\{G^{\rm M}_{r}, G^{\rm M}_{s}\right\}
=2L^{\rm M}_{r+s}+B^{\rm M}(r)\delta_{r+s}\,,
\quad\qquad\,\,\,\,\,\,&
\end{align}
where the anomalies are
\begin{align}
\,\,\,\,
\tilde{A}^{\rm M}(m)&=A^{\rm M}(m)=
\begin{cases}
\frac{D}{8}m^3
\qquad\qquad\quad\qquad\qquad\qquad
\qquad\qquad\qquad\qquad\qquad\,\,\,\,\,\,\,\
: \mbox{R sector}\,,
\\
\frac{D}{8}(m^3-m)
\\
\,\,\,\,\,\,\,
+\Big(\big[\hat{\mu}+\frac{1}{2}\big]^2
-[\hat\mu]^2-[\hat{\mu}]-2\big(\big[\hat{\mu}+\frac{1}{2}\big]
-[\hat{\mu}]-\frac{1}{2}\big)\hat\mu\Big) m
\,\,\,\,\,\,\,
: \mbox{NS sector}\,,\,\,\,
\end{cases}
\\
\tilde{B}^{\rm M}(r)&=B^{\rm M}(r)=
\begin{cases}
\frac{D}{2}r^2
\qquad\qquad\quad\qquad\qquad\qquad
\qquad\qquad\qquad\qquad\qquad\qquad\,\,\,
:\mbox{R sector}\,,
\\
\frac{D}{2}\big(r^2-\frac{1}{4}\big)
\\
\,\,\,\,\,\,
+\big[\hat{\mu}+\frac{1}{2}\big]^2
-[\hat\mu]^2-[\hat{\mu}]-2\big(\big[\hat{\mu}+\frac{1}{2}\big]
-[\hat{\mu}]-\frac{1}{2}\big)\hat\mu
\qquad\qquad
: \mbox{NS sector}\,.\,\,
\end{cases}
\end{align}

Finally, we simply comment on the super-Virasoro algebra of ghosts and antighosts.
\cite{G-S-W, Ohta-1}
The super-Virasoro generators of them are
\begin{align}
\tilde{L}^{\rm gh}_{n}&=\sum_{m\in \mathbb Z} (n+m) 
:\tilde{b}_{n-m} \tilde{c}_{m}:
+\frac{1}{2}\sum_{r\in \mathbb Z+ \varepsilon}\left(3n-2r\right)
:\tilde{\gamma}_{n-r}\tilde{\beta}_{r}:\,,\\
L^{\rm gh}_{n}&=\sum_{m\in \mathbb Z} (n+m) :b_{n-m} c_{m}:
+\frac{1}{2}\sum_{r\in \mathbb Z+ \varepsilon}\left(3n-2r\right)
:\gamma_{n-r}\beta_{r}: \,,\\
\tilde{G}^{\rm gh}_{r}&=-2\sum_{s\in \mathbb Z+\varepsilon}
\tilde{b}_{r-s}\tilde{\gamma}_{s}
+\frac{1}{2}\sum_{s\in \mathbb Z+\varepsilon}\left(s-3r\right)
\tilde{c}_{r-s}\tilde{\beta}_{s} \,,\\
G^{\rm gh}_{r}&=-2\sum_{s\in \mathbb Z+\varepsilon}
b_{r-s}\gamma_{s}
+\frac{1}{2}\sum_{s\in \mathbb Z+\varepsilon}\left(s-3r\right)
c_{r-s}\beta_{s}  \,,
\end{align}
where $\tilde{c}_{n}\,(c_{n})$ are the oscillator modes of the left (right)-moving
ghost of $c^{\alpha}$,
and $\tilde{b}_{n}\,(b_{n})$ are the oscillator modes of the left (right)-moving
antighost of $b_{\alpha \beta}$, which has only two components because
it is a traceless symmetric tensor, namely $b^{\,\,\,\alpha}_{\alpha}=0$.
\cite{Kato-Ogawa,Hwang}
Moreover, $\tilde{\gamma}_{r}\,(\gamma_{r})$ are the oscillator modes of 
the left (right)-moving ghost of $\gamma$ which has spinor components,
and $\tilde{\beta}_{r}\,(\beta_{r})$ are the oscillator modes of 
the left (right)-moving antighost of $\beta_{\alpha}$, which has only 
two components because it is traceless, namely 
$\gamma^{\alpha}\beta_{\alpha}=0$. (This $\gamma^{\alpha}$ is the
two-dimensional Dirac matrix.)
In terms of the modes the anticommutation relations are
$\{\tilde{c}_{m},\tilde{b}_{n}\}=\{c_{m},b_{n}\}=\delta_{m+n}$,
and the other anticommutators between 
$\tilde{c}_{n},\,\tilde{b}_{n},\,c_{n}$ and $b_{n}$ vanish.
The commutation relations are 
$[ \tilde{\gamma}_{r}, \tilde{\beta}_{s} ]
=[ \gamma_{r}, \beta_{s} ]=\delta_{r+s}$,
and the other commutators between 
$\tilde{\gamma}_{r}\,,\tilde{\beta}_{r}\,,\gamma_{r}$ and $\beta_{s}$ vanish.
Of course, the modes of $c^{\alpha}$ and $b_{\alpha\beta}$ are commutative
for the modes of $\gamma$ and $\beta_{\alpha}$.
It should be note that
the anomalies $\tilde{A}^{\rm gh}(m)$ and $A^{\rm gh}(m)$ of the algebras
$[\tilde{L}^{\rm gh}_{m}, \tilde{L}^{\rm gh}_{n}]$
and $[L^{\rm gh}_{m}, L^{\rm gh}_{n}]$ are $-\frac{5}{4}m^3$ in the R sector
and $-\frac{5}{4}m^3+\frac{1}{4}m$ in the NS sector.
In addition, the anomalies of $\tilde{B}^{\rm gh}(r)$ and $B^{\rm gh}(r)$ of
the algebras $\{\tilde{G}_{r}, \tilde{G}_{s}\}$ and 
$\{G_{r}, G_{s}\}$ are $-5r^{2}$ in the R sector and $-5r^{2}+\frac{1}{4}$
in the NS sector.


\section{The nilpotency of the BRST charge}

In the type II superstring theory, the left modes and the right modes are independent.
Therefore, the BRST charge can be decomposed into the left modes 
and the right modes as
\begin{align}
Q_{\rm B}=Q^{\rm L}_{\rm B}+Q^{\rm R}_{\rm B},
\end{align}
where
\begin{align}
Q_{\rm B}^{\rm L}
&=\sum_{m\in \mathbb{Z}}:\left[\tilde{L}_{m}^{\rm M}
+\frac{1}{2}\tilde{L}_{m}^{\rm gh}-a\delta_{m,0}\right]\tilde{c}_{-m}:
+\sum_{r\in \mathbb{Z}+\varepsilon}:\left[\tilde{G}^{\rm M}_{r}
+\frac{1}{2}\tilde{G}^{\rm gh}_{r}\right]\tilde{\gamma}_{-r}:\,\,,\\
Q_{\rm B}^{\rm R}
&=\sum_{m\in \mathbb{Z}}:\left[L_{m}^{\rm M}+\frac{1}{2}L_{m}^{\rm gh}
-a\delta_{m,0}\right]c_{-m}:
+\sum_{r\in \mathbb{Z}+\varepsilon}:\left[G^{\rm M}_{r}
+\frac{1}{2}G^{\rm gh}_{r}\right]\gamma_{-r}:\,.
\end{align}
Here $a$ is an ordering constant.
Concentrating our attention on the normal ordering of the mode operators, 
especially with regard to the ghosts, we obtain the square of $Q_{\rm B}$:
\begin{align}
Q_{\rm B}^{2}&=\frac{1}{2}
\Big[\left\{Q_{\rm B}^{\rm L},Q_{\rm B}^{\rm L}\right\}
+\left\{Q_{\rm B}^{\rm R},Q_{\rm B}^{\rm R}\right\}
\Big]\nonumber\\
&=\frac{1}{2}\sum_{m,n\in \mathbb{Z}}
\Big[\Big([\tilde{L}_{m},\tilde{L}_{n}]
-(m-n)\tilde{L}_{n+m}\Big)\tilde{c}_{-m}\tilde{c}_{-n}
+\Big([L_{m},L_{n}]-(m-n)L_{n+m}\Big)c_{-m}c_{-n}
\Big]\nonumber\\
&\,\,\,\,\,\,\,+\frac{1}{2}\sum_{r,s\in \mathbb{Z}+\varepsilon}
\Big[\Big(\{\tilde{G}_{r}, \tilde{G}_{s}\}
-2\tilde{L}_{r+s}\Big)\tilde{\gamma}_{-r}\tilde{\gamma}_{-s}
+\Big(\left\{G_{r}, G_{s}\right\}
-2L_{r+s}\Big)\gamma_{-r}\gamma_{-s}\Big]\nonumber\\
&=\frac{1}{2}\sum_{m\in \mathbb{Z}}A(m)\left(\tilde{c}_{-m}\tilde{c}_{m}
+c_{-m}c_{m}\right)
+\frac{1}{2}\sum_{m\in \mathbb{Z}+\varepsilon}B(r)
\left(\tilde{\gamma}_{-r}\tilde{\gamma}_{r}
+\gamma_{-r}\gamma_{r}\right)\,,
\end{align}
where $\tilde{L}_{m}$, $L_{m}$, $\tilde{G}_{r}$ and $G_{r}$ are
the total super-Virasoro generators as follows:
\begin{align}
\tilde{L}_{m}&=\tilde{L}_{m}^{\rm M}+\tilde{L}_{m}^{\rm gh}-a\delta_{m,0}\,,
\qquad L_{m}=L_{m}^{\rm M}+L_{m}^{\rm gh}-a\delta_{m,0}\,,\\
\tilde{G}_{r}&=\tilde{G}^{\rm M}_{r}+\tilde{G}^{\rm gh}_{r}\,,
\qquad\qquad\quad\,\,\, G_{r}=G^{\rm M}_{r}+G^{\rm gh}_{r}\,.
\end{align}
$A(m)$ and $B(r)$ are the total anomalies from the total super-Virasoro
generators, and we can exactly demonstrate them using the results of 
the anomalies of the matter in the previous section and 
the known anomalies of the ghosts.
As a result, the total anomalies are
\begin{align}
A(m)=&
\begin{cases}
\frac{D-10}{8}m^{3}+2am
\qquad\qquad\qquad\qquad\qquad\qquad\
\qquad\qquad\qquad\qquad\,\,\,\,\,\,\,\,\,\,
: \mbox{R sector}\,,
\\
\frac{D-10}{8}(m^{3}-m)\\
\quad\,\,\,\,
+2\Big(a-\frac{1}{2}+\big[\hat{\mu}+\frac{1}{2}\big]^2
-[\hat\mu]^2-[\hat{\mu}]-2\big(\big[\hat{\mu}+\frac{1}{2}\big]
-[\hat{\mu}]-\frac{1}{2}\big)\hat\mu
\Big) m
\,\,\,
: \mbox{NS sector}\,,
\end{cases}
\\
B(r)=&
\begin{cases}
\frac{D-10}{2}r^{2}+2a
\qquad\qquad\quad\qquad\qquad\qquad\qquad\
\qquad\qquad\quad\qquad\quad\qquad
: \mbox{R sector}\,,
\\
\frac{D-10}{2}\big(r^{2}-\frac{1}{4}\big)
\\
\quad\,\,\,\,
+2\Big(a-\frac{1}{2}+\big[\hat{\mu}+\frac{1}{2}\big]^2
-[\hat\mu]^2-[\hat{\mu}]-2\big(\big[\hat{\mu}+\frac{1}{2}\big]
-[\hat{\mu}]-\frac{1}{2}\big)\hat\mu
\Big)
\,\,\,\,\,\,\,\,
: \mbox{NS sector}\,.
\end{cases}
\label{anomaly}
\end{align}

If the anomalies are zero, the square of the BRST charge vanishes.
Because the BRST charge must have the property of nilpotency, 
the anomalies must be zero.
Thus we can determine the number of spacetime dimensions 
and the ordering constant in the $\it{pp}$-wave background with the NS-NS
flux:
\begin{align}
&\,\,\,\qquad 
D=10,\quad a=0\quad\qquad\qquad\qquad\qquad\qquad\qquad\
\qquad\quad\qquad\qquad\qquad\quad\,\,
 :\mbox{R sector}\,,\\
&\,\,\,\qquad 
D=10,\quad a=\frac{1}{2}-\big[\hat{\mu}+\frac{1}{2}\big]^2
+[\hat\mu]^2+[\hat{\mu}]+2\big(\big[\hat{\mu}+\frac{1}{2}\big]
-[\hat{\mu}]-\frac{1}{2}\big)\hat\mu\qquad\,\,
 :\mbox{NS sector}\,.
\end{align}

Considering the spectrum of the superstring 
in the {\it pp}-wave background,
the physical state must satisfy $Q_{\rm B}|{\rm phys}\rangle=0$.
Using our free-modes, we can get them.


\section{Conclusion}

In this paper we have canonically quantized the closed RNS superstring in 
the {\it pp}-wave background with the non-zero flux of the $B_{\mu\nu}$ field 
using the covariant BRST operator formalism. 
In this {\it pp}-wave background with the flux,  we have constructed
the general operator solutions and the free-mode representations 
of all the covariant string coordinates and fermions.
Moreover, we proved that the free-mode representations
satisfy both the equal-time canonical (anti)commutation relations 
between all the covariant superstring fields and the Heisenberg 
equations of motion, whose form is the same as that of the Euler-Lagrange 
equations of motion in the {\it pp}-wave background with the flux.
It is worth noting that the zero mode $x^{-}$ of $X^{-}_{0}$
has played important roles in this study.
Since the energy-momentum tensor and the supercurrent take 
very simple forms in the free-mode representations 
of the covariant string coordinates and fermions,
we have been able to calculate the anomaly in the super-Virasoro algebra.
Using this anomaly, we have determined the number of dimensions of 
spacetime and the ordering constant from the nilpotency condition of 
the BRST charge in the {\it pp}-wave background with the flux.
The spacetime supersymmetry is realized
due to the condition like the GSO projection based on 
the difference between the R sector and the NS sector.


\end{document}